\documentclass{article}
\usepackage[utf8]{inputenc}
\usepackage{csquotes}
\usepackage[english]{babel}
\usepackage{url}
\usepackage{microtype}

\usepackage{authblk}
\usepackage[style=apa, natbib=true, backend=biber]{biblatex}
\usepackage{adjustbox}
\usepackage{amsmath}
\usepackage{rotating}
\usepackage{tablefootnote}
\usepackage{changepage}
\usepackage{dirtytalk}
\usepackage{caption} 
\usepackage{multirow}
\usepackage{array}
\usepackage{tabularx}
\usepackage{wrapfig}
\usepackage[colorinlistoftodos]{todonotes}

\newcommand{\beginsupplement}{%
        \setcounter{table}{0}
        \renewcommand{\thetable}{S\arabic{table}}%
        \setcounter{figure}{0}
        \renewcommand{\thefigure}{S\arabic{figure}}%
     }

\title{A framework for model-assisted T $\times$ E $\times$ M exploration in maize}

\author[1]{Jennifer Hsiao}
\author[3]{Soo-Hyung Kim}
\author[4]{Dennis J. Timlin}
\author[5]{Nathaniel D. Mueller}
\author[1,2]{Abigail L.S. Swann}
\affil[1]{University of Washington, Department of Biology, Seattle, WA, United States}
\affil[2]{University of Washington, Department of Biology, Department of Atmospheric Sciences, Seattle, WA, United States}
\affil[3]{University of Washington, School of Environmental and Forest Sciences, Seattle, WA, United States}
\affil[4]{USDA-ARS Adaptive Cropping Systems Laboratory, Beltsville, MD, United States}
\affil[5]{Colorado State University, Department of Ecosystem Science and Sustainability and Department of Soil and Crop Sciences, Fort Collins, CO, United States}

\begin{document}

\maketitle

\begin{abstract}
Breeding for new crop characteristics and adjusting management practices are critical avenues to mitigate yield loss and maintain yield stability under a changing climate. However, identifying high-performing plant traits and management options for different growing regions through traditional breeding practices and agronomic field trials is often time and resource-intensive. Mechanistic crop simulation models can serve as powerful tools to help synthesize cropping information, set breeding targets, and develop adaptation strategies to sustain food production. In this study, we develop a modeling framework for a mechanistic crop model (MAIZSIM) to run many simulations within a trait $\times$ environment $\times$ management landscape and demonstrate how such a modeling framework could be used to identify ideal trait-management combinations that maximize yield and yield stability for different agro-climate regions in the US. 
\end{abstract}

\section{Introduction}
Food demand is increasing but our ability to sustain crop productivity will be impacted by a warming climate. Breeding has consistently played a critical role in the progress of continuous yield gain and was estimated to account for up to 50-60\% of the total on-farm yield gain in the past several decades \citep{Duvick2005}. These gains through genetic improvements are complemented by changes in management practices, such as an increase in fertilizer use, chemical weed control, higher planting densities, and earlier planting dates \citep{Kucharik2008, Cardwell1982}. Recently, however, practices such as nitrogen application and weed control are nearly fully exploited in the US corn belt; simple adjustments in management strategies alone are likely insufficient to sustain an increasing yield trend. Additional yield gains would need to rely further on genetic improvements in new cultivars, as well as management changes that accompany climate-resilient characteristics to fully leverage the interactions among genetics, environment, and management (the G $\times$ E $\times$ M paradigm, \cite{Hatfield2015}).

Continued development of new cultivars better-suited for future climate is critical for sustaining current yield trends or to prevent yield loss \citep{Challinor2017,Burke2009}. Progress in breeding for climate adaptation has been demonstrated in several areas, including changes in morphological traits (e.g. improved root system architecture that improves soil water access; \cite{Hammer2009}), increases in drought and salinity tolerance \citep{Fita2015, Messina2020}, improvements in physiological traits (e.g. greater nitrogen use efficiency; \cite{Fischer2010}), and shifts in copping duration \citep{Zhu2018}, to name a few. Our ability to utilize the genetic diversity preserved in wild relatives, landrace species, and undomesticated wild species to develop new climate-ready cultivars is increasingly important to achieve sustainable and intensified food production \citep{McCouch2013, Godfray2010}. 

Maize trait changes in the past few decades have also been accompanied by shifts in crop management practices, such as more erect plant forms that facilitated notable increases in planting densities \citep{Duvick2005}, shifts towards earlier planting dates by about 3 days per decade \citep{Butler2018, Zhu2018}, increases in nutrient supply \citep{Duvick2005}, and increases in area irrigated \citep{Mueller:2015es}. In addition, the suitability of a cultivar often varies considerably across environmental gradients (e.g. \cite{Messina2015}), thus optimal plant traits and management options are usually identified within defined target environments \citep{Cooper2016}. This breeding strategy allows for designing different cultivars to perform favorably and withstand stresses in their target environments, and what is considered \say{ideal} may differ between locations and climate. We expect optimal management to shift under future climate conditions and in combination with different phenotypic traits, providing important means of adaptation in many systems \citep{Deryng2011}. 

Mechanistic modeling tools that integrate physiological, morphological, and phenological properties of a crop (\emph{G}), their performance under different management options (\emph{M}), and their interactions with the surrounding environment (\emph{E}) on a whole-plant level can serve as useful tools for breeding practices through the quantification of a yield-trait-performance landscape \citep{Messina2011a}. The structure of such models allow for testing effects of traits (e.g. leaf elongation rate, total leaf number) on integrated outcomes such as yield. While mechanistic crop models may not specifically describe genetic-level properties, higher-level traits are often used as proxies to describe the underlying genotype. This makes models ideal tools to test and screen for potentially promising traits and management (\emph{T $\times$ M}) combinations under different climate and environmental conditions (\emph{E}) as a first step before carrying out actual breeding practices \citep{Andrivon2012, Messina2011a}, and on large scales that are often not feasible under actual experimental settings \citep{Peng2020, Cooper2020, Hammer2020}. Such information can further be used to synthesize cropping knowledge, set breeding targets, and develop climate-targeted adaptation strategies to sustain food production. 

Despite broad recognition that mechanistic, process-based crop simulation models can be a powerful tool to synthesize cropping information, set breeding targets for developing climate-ready crops, and develop adaptation strategies for sustaining food production \citep{Muller2019,Messina2020}, few comprehensive studies have been performed to produce climate-specific trait and management combinations for staple crops including maize in the US, a necessity given rapidly changing environmental conditions facing the US cropping systems. In this study, we construct a modeling framework to identify targeted plant traits and effective crop management to achieve maximum crop performance in both the current and future climates. Specifically, we addressed how an ensemble of plant traits (i.e. physiology, morphology, phenology) combined with realistic adjustments to management choices (i.e. shifting planting dates, planting density, and row spacing) can be used to build resilience and improve productivity under the stresses induced from a changing climate.

\section{Material and methods}
\begin{figure}[htp]
        \centering
        \includegraphics[width=10cm]{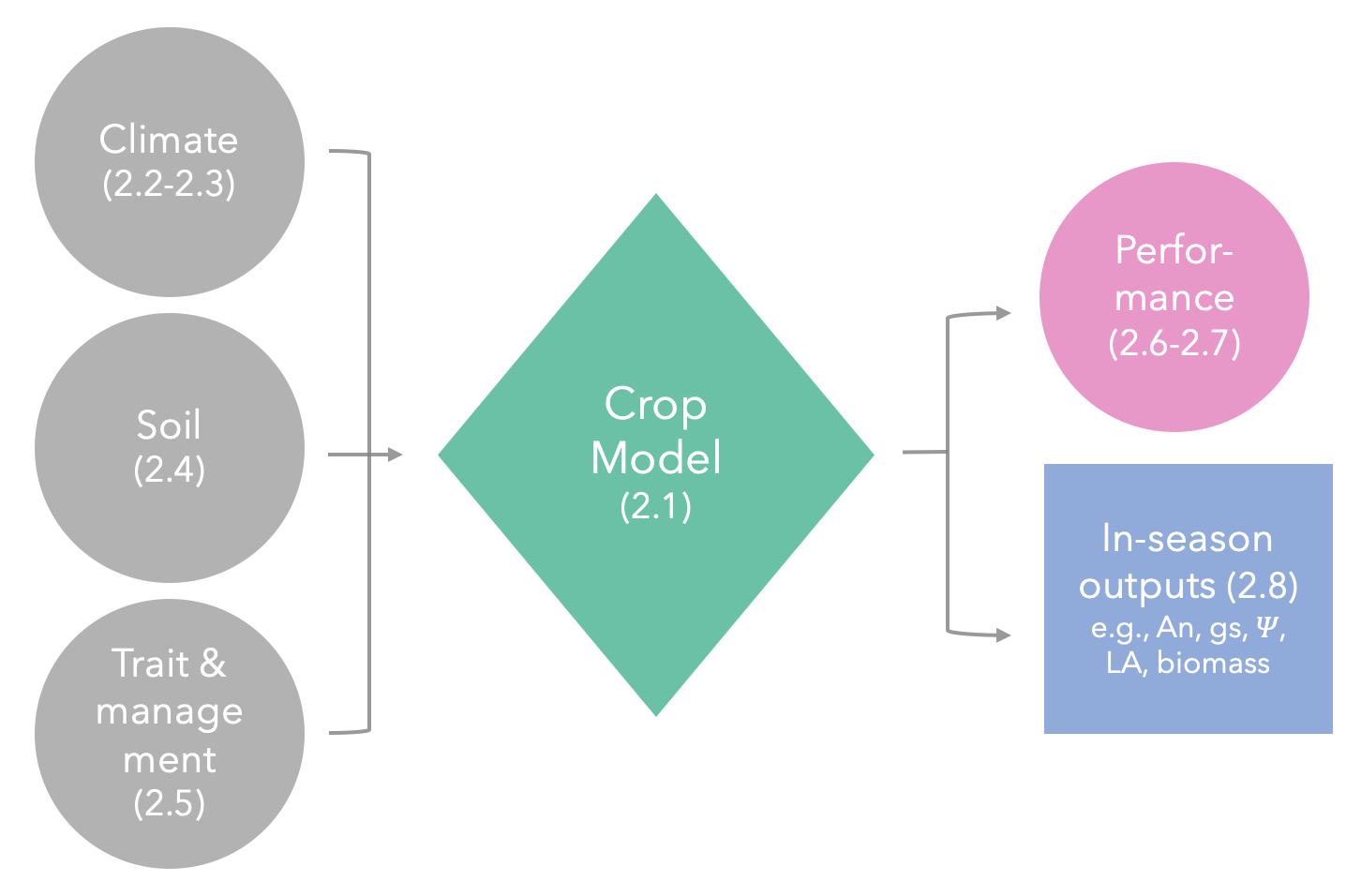}
        \caption{Diagram of the data-model framework.}
        \label{fig:framework}
\end{figure}

We set up a data-model framework to quantitatively identify high performing regions within a T $\times$ E $\times$ M landscape. The framework consists of three main components (Fig. \ref{fig:framework}): \emph{1)} a process-based crop simulation model (section \ref{methods_maizsim}), \emph{2)} model inputs to drive the model, including present-day climate information (section \ref{methods_present_climate}), idealized future climate information (section \ref{methods_future_climate}), simulation site soil information (section \ref{methods_soil}), and sampled trait and management options (section \ref{methods_params}), and \emph{3)} processed model outputs that identify performance within the T $\times$ E $\times$ M landscape (section \ref{methods_performance} - \ref{methods_performance_loc}), and summarized in-season growth outputs (section \ref{methods_inseason}).

\subsection{Process-based crop simulation model - MAIZSIM} \label{methods_maizsim}
MAIZSIM is a deterministic and dynamic model developed and calibrated for maize plants to represent key physiological and physical processes such as gas exchange, canopy radiative transfer, carbon partitioning, water relations, nitrogen dynamics and phenology \citep{Kim_Modeling_2012}. MAIZSIM interfaces with a 2-dimensional finite element model (2DSOIL) that simulates a dynamic soil water and nutrient vertical 2D profile \citep{Timlin:1996gk}. The coupled model responds to daily or hourly meteorological information throughout the growing season that includes temperature, relative humidity, solar radiation, and CO\textsubscript{2} concentrations.

At the leaf-level, MAIZSIM captures gas exchange processes through a C4 photosynthesis model \citep{VonCaemmerer} coupled with a stomatal conductance model \citep{BBW} and an energy balance equation \citep{Collatz:1992ks}; leaf-level gas exchange processes are scaled to canopy-levels using a sunlit/shaded leaf framework \citep{dePury:1997uu}. The model simulates crop development throughout the growing season following a nonlinear temperature response \citep{Yin1995}, and adopts a leaf area model developed by \cite{Lizaso2003} to describe the expansion and senescence of individual leaves. MAIZSIM dynamically simulates leaf water potential and uses it to trigger water stress responses such as reduced growth rate and hastened senescence when values drop below designated thresholds \citep{Yang:2009cm}. 

The model has been validated at different scales – including physiological aspects such as gas exchange \citep{YYang:2009eu}, leaf development and biomass gain \citep{Kim_Modeling_2012}, leaf growth water stress responses \citep{Yang:2009cm}, as well as field-level validations in AgMIP projects \citep{Bassu_How_2014,Kimball2019} and FACE site studies \citep{Durand_How_2017} that tested for yield responses to different temperature and CO\textsubscript{2} conditions. The model has also recently been used to test the independent impacts of temperature versus VPD on growth and yield in maize growing regions in the US \citep{Hsiao2019}.

\begin{figure}[htp]
        \centering
        \includegraphics[width=11cm]{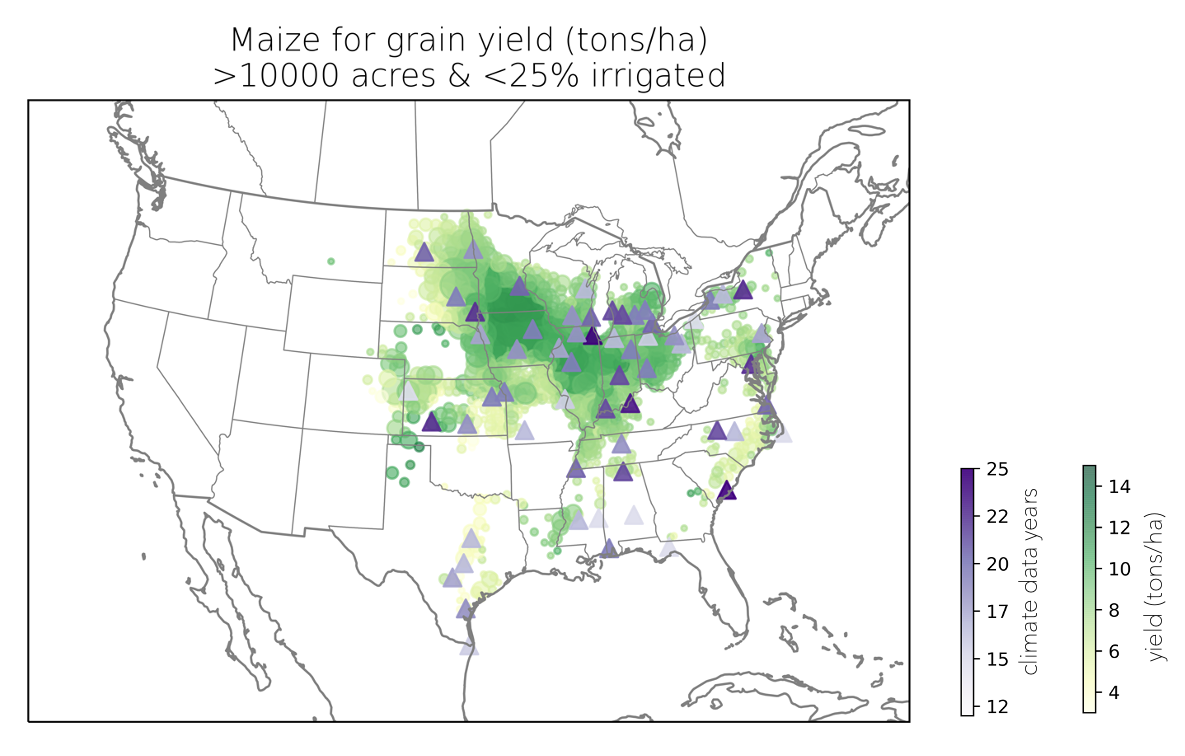}
        \caption{Simulation sites and number of years simulated (purple triangles), along with historic maize yield and planting area data (green circles). Colors indicate yield and circle size indicate planting area.}
        \label{fig:map_simsites}
\end{figure}

\subsection{Present-day climate data} \label{methods_present_climate}
We assembled hourly data of temperature, relative humidity, precipitation, and solar radiation over years 1961-2005 for our simulation sites as weather data input for our model simulations. Specifically, we accessed hourly air temperature ($T_{air}$), dew point temperature ($T_{dew}$), and precipitation data from the NOAA National Center for Environmental Information Integrated Surface Hourly database (\url{https://www.ncdc.noaa.gov/isd}), and hourly solar radiation data from the National Solar Radiation Data Base (\url{https://nsrdb.nrel.gov/data-sets/archives.html}).

We followed the Clausius-Clapeyron equation (Eqn. \ref{eq:Es}-\ref{eq:E}) to back out atmospheric humidity information in the form of relative humidity (RH) from $T_{air}$ and $T_{dew}$:

\begin{equation}\label{eq:Es}
E_s=E_{sref}\cdot e^{\frac{L_v}{R_v}\cdot(\frac{1}{T_{ref}}-\frac{1}{T_{air}})}
\end{equation}
\begin{equation}\label{eq:E}
E=E_{sref}\cdot e^{\frac{L_v}{R_v}\cdot(\frac{1}{T_{ref}}-\frac{1}{T_{dew}})}
\end{equation}
The equation uses the saturation vapor pressure ($E_{sref}$, 6.11 mb) at a reference temperature ($T_{ref}$, 273.15 K), the vaporization latent heat ($L_v$, 2.5$\cdot$10\textsuperscript{6} J kg\textsuperscript{-1}), and the gas constant ($R_v$, 461 J K\textsuperscript{-1}kg\textsuperscript{-1}) to calculate the saturated water vapor pressure ($E_s$, mb) and the actual water vapor pressure ($E$, mb) at air temperature ($T_{air}$, K). We then use $E$ and $E_s$ to calculate $RH$ (\%) (Eqn. \ref{eq:RH}):
\begin{equation}\label{eq:RH}
RH=\frac{E}{E_s}
\end{equation}

\begin{figure}[ht]
        \centering
        \includegraphics[width=11cm]{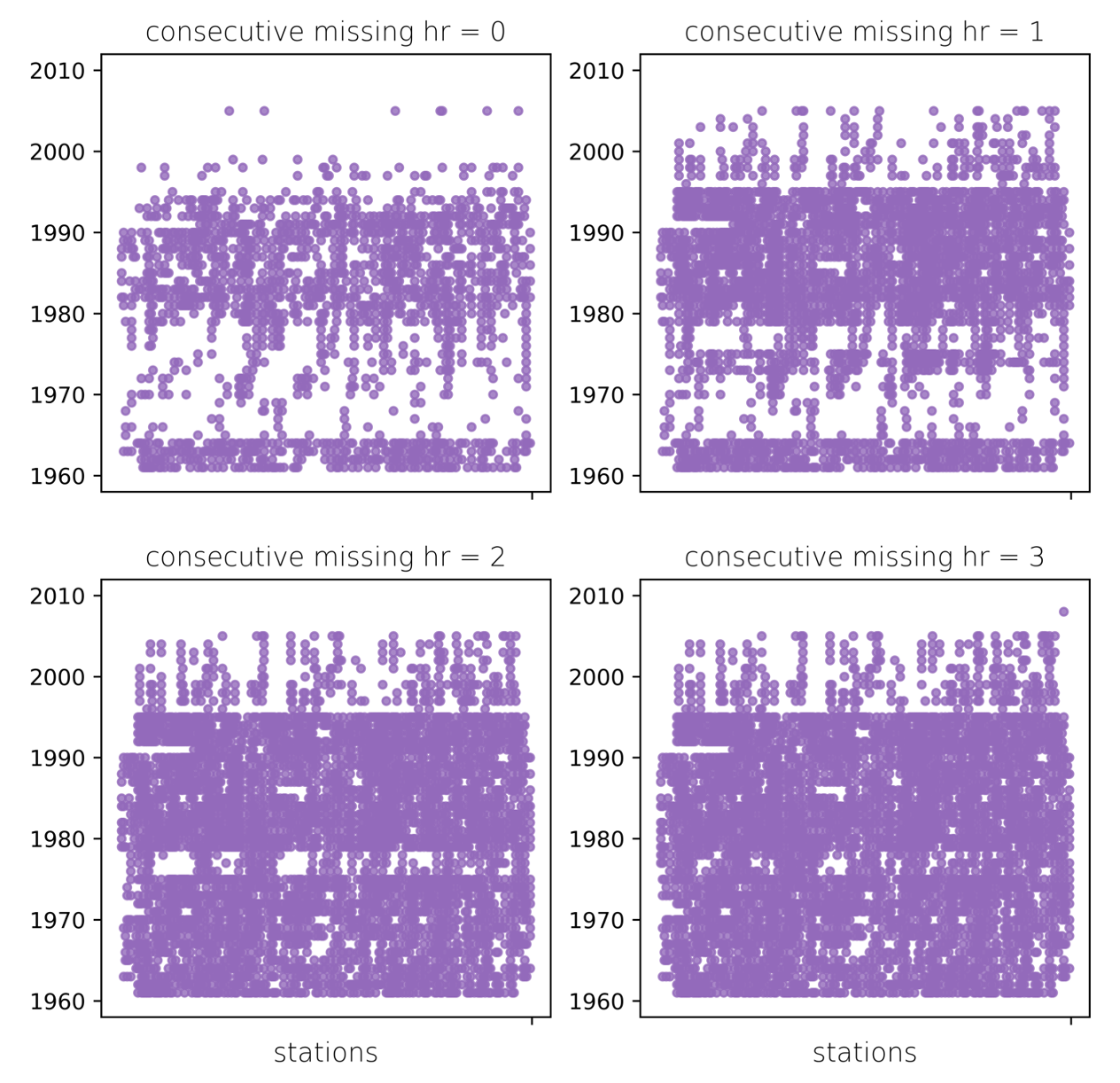}
        \caption{Available weather data based on different gap-filling intervals. For example, if consecutive missing hours equals 0, then only site-years with complete hourly weather data records will be included for weather data. If consecutive missing hour equals 1, site-years with gaps no greater than one hour consecutively will be included and gap-filled linearly.}
        \label{fig:scatter_weadata}
\end{figure}

\begin{figure}[htp]
        \centering
        \includegraphics[width=8cm]{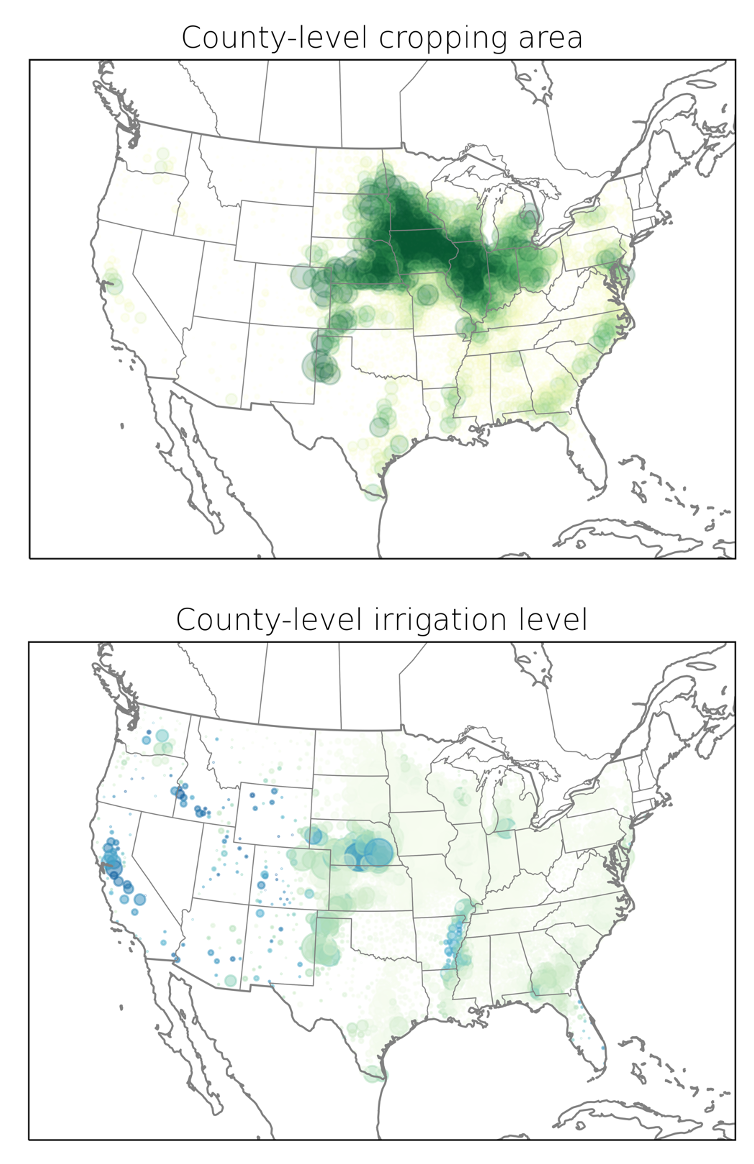}
        \caption{Maize planting area (top) and irrigation levels (bottom) across continental U.S.}
        \label{fig:maps_plantarea_irri}
\end{figure}

We selected overlapping sites and years that had data available from both the Integrated Surface Hourly Data Base and the National Solar Radiation Data Base over years 1961-2005 and filtered for site-years that had less than two consecutive hours of missing data throughout the growing season (broadly defined to be between February 1\textsuperscript{st} – November 30\textsuperscript{th}) and retained at least two-thirds of the weather data (Fig. \ref{fig:scatter_weadata}). We then gap-filled any missing data by linearly interpolating the missing information with weather data of the hours prior and post the missing data point. Next, we linked valid weather stations with maize planting area and irrigation level data accessed through the United States Department of Agriculture – National Agriculture Statistics Service (USDA- NASS, \url{https://www.nass.usda.gov/Data_and_Statistics/index.php}). Specifically, we calculated the average maize planting area across our simulation period (1961-2005) in the continental US and accessed average irrigation level (\%) for the same sites through four available census years (1997, 2002, 2007, 2012) (Fig. \ref{fig:maps_plantarea_irri}). We used the planting area and irrigation level averaged across five USDA-NASS sites closest to each weather station (via Euclidean distance) to represent their cropping information, and to exclude sites that either had less than 10,000 acres of corn planted or had greater than 25 \% of crop land irrigated. We excluded sites with less than 15 years of data to insure sufficient sampling to assess inter-annual climate variability \citep{Soltani2003,VanWart2013}. Following this method, we were able to compile 1160 site-years of meteorology data for our simulations, which included a total of 60 sites, each site with available weather data ranging from 15-27 years (Fig. \ref{fig:map_simsites}).

\subsection{Idealized projected climate} \label{methods_future_climate}
\begin{figure}[htp]
        \centering
        \includegraphics[width=12cm]{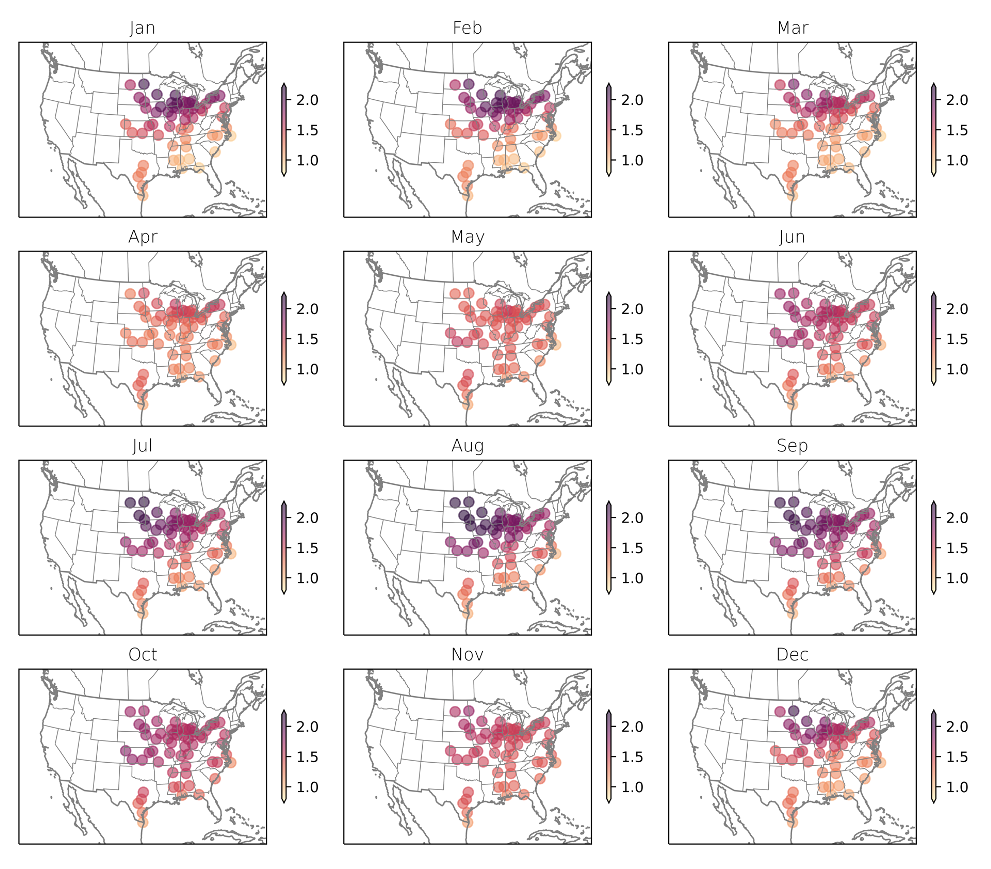}
        \caption{Monthly pattern of warming derived from CMIP6 multimodel means for our simulation sites. Numbers in color bar indicate temperature scaling values to multiply with global average climate sensitivity to calculate projected warming for each simulation site. For example, under our assumption of 3.1\textdegree C global average warming by 2100, a scaling value of 2 for a specific simulation site will equal a total warming of 6.2\textdegree C for that location.}
        \label{fig:maps_temp_scaling}
\end{figure}

We assembled idealized projected climate information at two future time points, 2050 and 2100, to analyze crop performance shifts under future climate (Table \ref{table:climate_treatment}). Specifically, we created monthly temperature and relative humidity anomaly maps under a substantial but not extreme greenhouse gas emissions scenario (SSP3-7.0, \citeauthor{Riahi2017}, \citeyear{Riahi2017}) from the latest Coupled Model Intercomparison Project version 6 (CMIP6) outputs; we used these anomaly maps to calculate location-specific warming and associated changes in relative humidity levels throughout the growing season for each simulation site (Fig. \ref{fig:maps_temp_scaling}). This method preserves correlations between climate variables (i.e., between temperature, relative humidity, and solar radiation) on short timescales and limits known biases in modeled variability \citep{VargasZeppetello2019, Donat2017}. Since both magnitude and pattern of future precipitation projections are highly uncertain, we applied a general trend of precipitation reduction in accordance to the SSP3-7.0 scenario, and increased atmospheric concentrations of CO$_2$ to ~550 ppm and ~850 ppm for years 2050 and 2100, respectively \citep{ONeill2016}.

\begin{table}[h]
        \centering
        \caption{Description of idealized climate treatments with projected changes in temperature (T), relative humidity (RH), precipitation (precip.), and projected CO\textsubscript{2} concentrations by years 2050 and 2100 under the SSP3-7.0 emission scenario.}
        \label{table:climate_treatment}
        \begin{tabular}{ll}
        \hline
        Year & Climate Scenario \\
        \hline
        2050 & + 1.4 \textdegree C mean T, -RH, -15\% precip, 550 ppm \\
        2100 & + 3.1 \textdegree C mean T, -RH, -30\% precip, 850 ppm \\
        \hline
        \end{tabular}
\end{table}

\subsection{Soil data} \label{methods_soil}
We used soil information from USDA-NASS locations nearby our simulation sites to curate site-specific soil files for each location. Soil properties are highly heterogeneous, and since our simulation sites are based on weather station locations that do not directly come from agricultural land, we use this method to broadly represent soil makeup of agricultural sites within the region without skewing towards any particular site. We queried soil information from the National Resources of Conservation Services (NRCS) SSURGO soil database \citep{ssurgo} to identify soil properties for each NASS location with maize planting area greater than 10000 acres and irrigation levels less than 25\%. For each site, we accessed soil information at five depth categories (surface, 50, 100, 150, and 200 cm), which included sand--silt--clay--organic matter composition, soil bulk density (the oven dry weight of less than 2 mm soil material per unit volume of soil at a water tension of 1/3 bar), and the volumetric content of soil water retained at a tension of 1/3 bar (33 kPa, field capacity) and 15 bar (1500 kPa, wilting point) expressed as a percentage of the whole soil. With the sand--silt--clay composition, we categorized the queried soil data into 12 texture groups following the USDA Textural Soil Classification \citep{soilclassification} and excluded sites classified as Sandy or Clay due to their lack of representation in agricultural fields. Next, we determined the soil class within each depth category for all our simulation sites by assigning it the most prevalent soil class from it’s 11 nearest NASS sites calculated through Euclidean distance, and assigned it the mean soil conditions of that texture-depth class averaged across all NASS sites within that category. Finally, we estimated soil hydraulic properties of each soil type through a water release curve predicted by the van Genuchten equation \citep{vanGenuchten}.

\begin{figure}[htp]
        \centering
        \includegraphics[width=10cm]{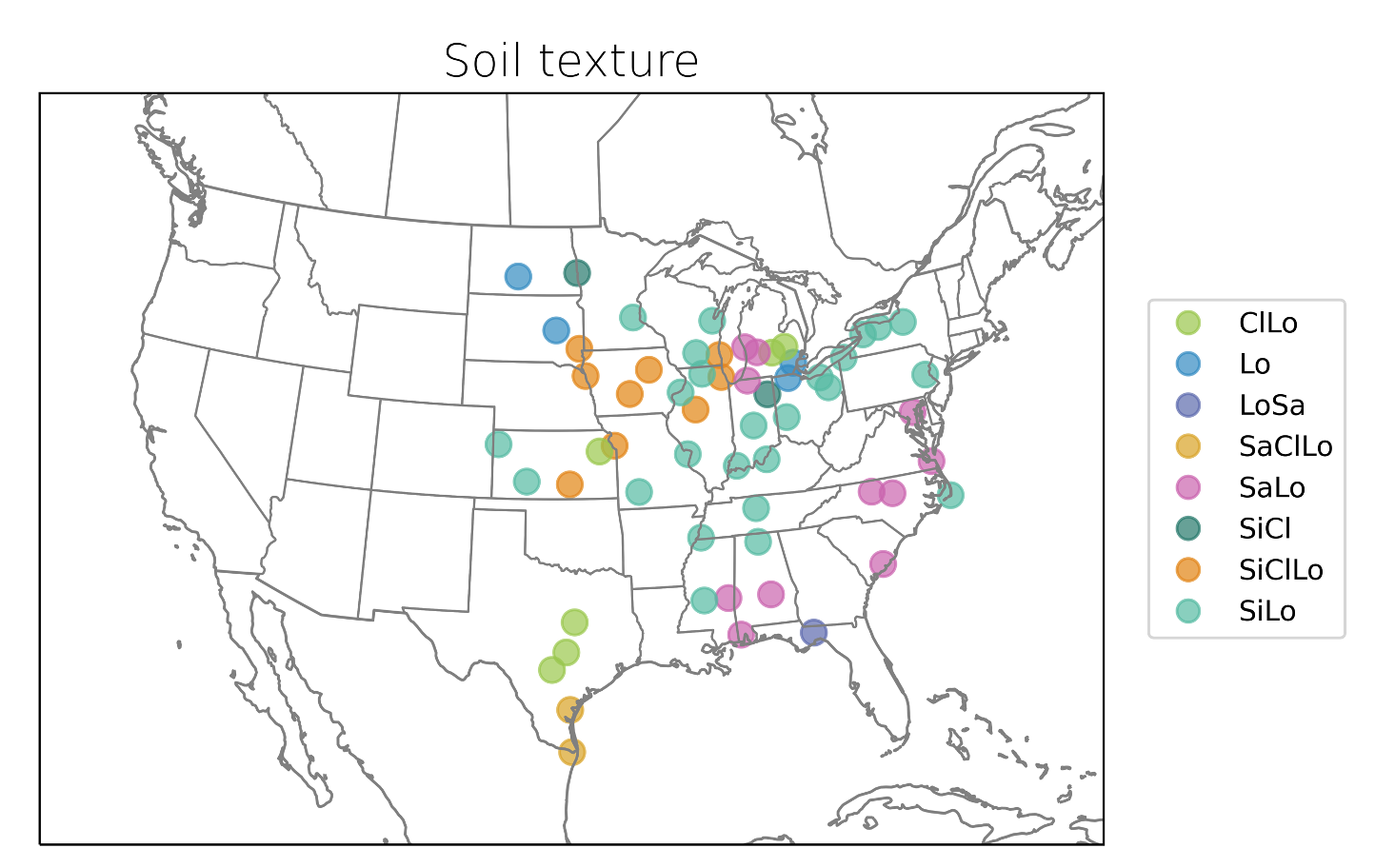}
        \caption{Soil texture across simulation sites. Soil texture categories include clay loam (ClLo), loam (Lo), loamy sand (LoSa), sandy clay loam (SaLoLo), sandy loam (SaLo), silty clay (SiCl), silty clay loam (SiClLo), and silty loam (SiLo).}
        \label{fig:map_soils_sitetexture}
\end{figure}

\subsection{Sampling within the trait and management space} \label{methods_params}
We selected several key model parameters that represent a range of maize traits and management options to investigate combinations that lead to high performance under present and future climate conditions. Since we do not have robust observation-based data on the natural distribution and boundaries of most parameters, we assumed a uniform distribution and set biologically reasonable boundaries around literature-based default values (Table \ref{table:params}). We assumed all parameters to be non-correlated and used a Latin hypercube sampling method \citep{McKay1979} to create 100 different trait-management (T $\times$ M) combinations within the parameter space.

\begin{sidewaystable}
        \centering
        \caption{MAIZSIM parameters tested for yield optimization}
        \label{table:params}
        \begin{tabular}{p{2cm}p{1.5cm}p{5cm}p{2.5cm}p{5cm}}
        
        \hline
        Processes & Params & Description & Default (Range) & Citation \\
        \hline
        \\
        Physiology & g\textsubscript{1} & Ball-Berry g\textsubscript{s} model slope & 4 (2$\sim$6) & \cite{Miner2017,Shekoofa2016} \\
        & V\textsubscript{cmax} & Max RUBISCO capacity & 65 (65$\sim$80) & \cite{Kim:2006dd, Wu2019} \\
        & J\textsubscript{max} & Max electron transport rate & 350 (350$\sim$420) & \cite{Kim:2006dd,Wu2019} \\
        & phyf & Reference leaf water potential (MPa) used to describe stomata sensitivity to leaf water potential & -1.9 (-3$\sim$-1) & \cite{Tuzet:2003td,YYang:2009eu,Shekoofa2016} \\
        Phenology & SG & Duration that leaves maintain active function (stay-green) after reaching maturity & 3 (2$\sim$6) & \cite{Zhang2019,Gregersen2013,TIMLIN201957} \\
        & gleaf & Total leaf number & 19 (11$\sim$25) & \cite{Parent2018,Padilla2005} \\
        & LTAR & Max leaf tip appearance rate (leaves/day) & 0.5 (0.4$\sim$0.8) & \cite{Kim_Modeling_2012,Padilla2005} \\
        Morphology & LM & Leaf length of the longest leaf (cm) & 115 (80$\sim$120) &  \\
        & LAF & Leaf angle factor & 1.37 (0.9$\sim$1.4) & \cite{Campbell1986,Dzievit2019} \\
        Management & gdd & Growing degree days accumulated by sowing & 100 (80$\sim$160) & \cite{Sacks_Crop_2011,TIMLIN201957} \\
        & pop & Density (plants/m\textsuperscript{2}) & 8 (6$\sim$10) & \cite{TIMLIN201957,Stone2000} \\

        \hline
        \end{tabular}
\end{sidewaystable}

\subsection{Performance within the T $\times$ E $\times$ M landscape}  \label{methods_performance}
We defined high crop performance as crops that achieve high yield (i.e. yield mean across years) and high yield stability (i.e. yield dispersion across years). We developed a cost function (Eqn. \ref{eq:COSTFUNC}) to quantify the performance of any T $\times$ M combination by calculating its distance to a theoretical best-performing combination within the yield and yield stability space (Eqn. \ref{eq:COSTFUNC}):

\begin{equation}\label{eq:COSTFUNC}
        D_{score} = \sqrt{w_{yield}*(y_{mean}-y_{max})^2 + w_{disp}*(y_{disp}-d_{min})^2}
\end{equation}
        
\noindent $y_{mean}$ and $y_{disp}$ represent mean yield and yield dispersion (variance/mean) across years for the target T $\times$ M combination at a specific simulation site, respectively. We standardized yield and dispersion to values between 0 and 1 to avoid skewed contribution due to difference in scale. $y_{max}$ and $d_{min}$ denote the standardized maximum mean yield (1) and minimum yield dispersion (i.e., maximum yield stability, 0) achieved within all T $\times$ M combinations at a specific simulation site. $w_{yield}$ and $w_{disp}$ are empirical coefficients between 0 and 1 that assign weighted importance to yield mean and yield dispersion, respectively.

\begin{figure}[htp]
        \centering
        \includegraphics[width=12cm]{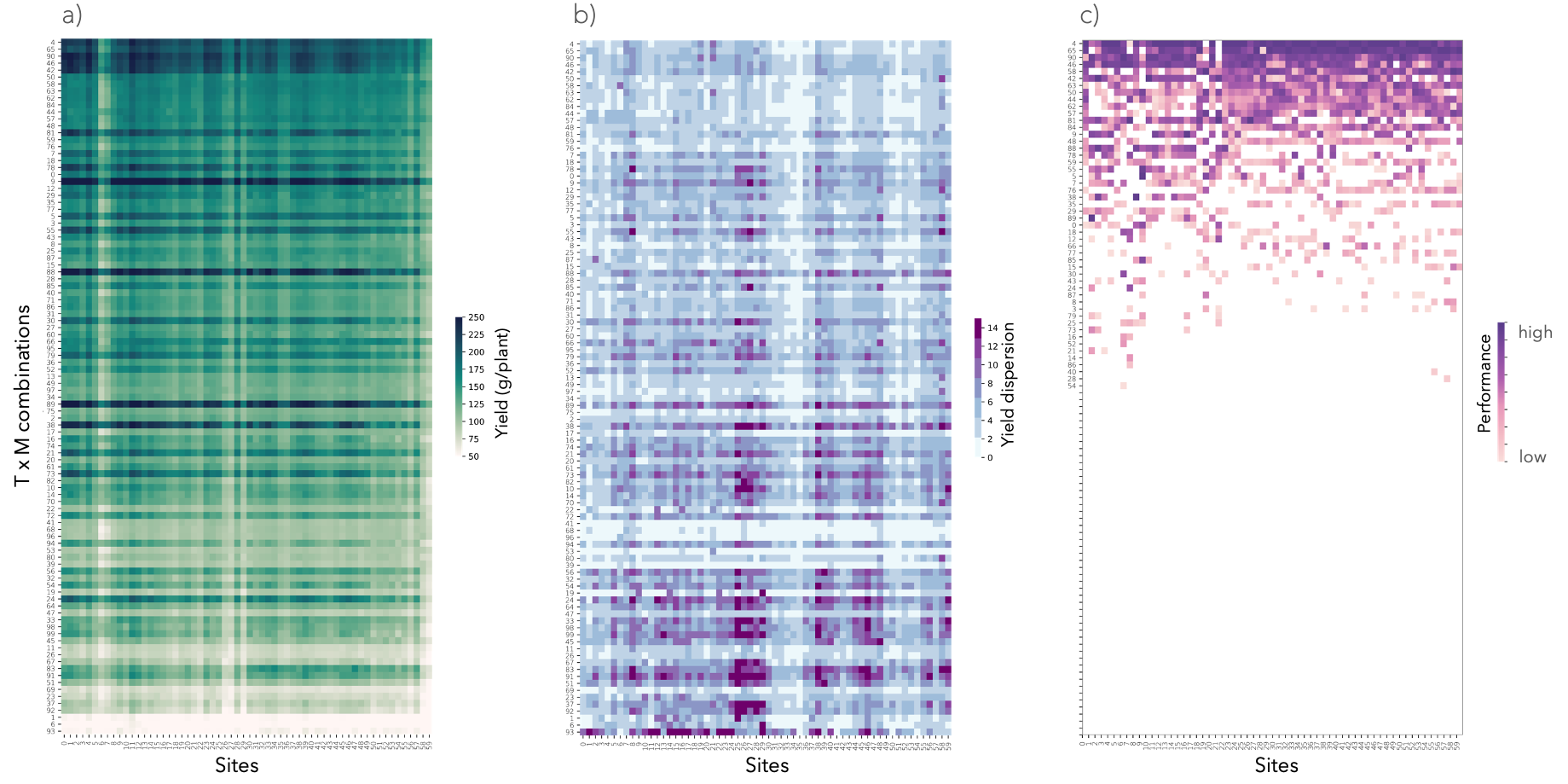}
        \caption{Simulated a) yield, b) yield dispersion, and c) performance ranking of across T $\times$ M combinations and locations. T $\times$ M combinations are ordered from top to bottom starting from the highest performance ranking. Sites are order from left to right from sites located in the south to the north.}
        \label{fig:performance}
\end{figure}

We used the calculated $D_{score}$ to rank the top 20 performing T $\times$ M combinations for each simulation site. We determined an overall ranking for each T $\times$ M combination based on their rankings across all simulation sites (Fig. \ref{fig:performance}a). With this method, T $\times$ M combinations with high rankings across a few sites versus combinations with medium ranking across several sites can all result in overall high performance. T $\times$ M combinations that do not rank within the top 20 performers at any site will not receive a ranking.

\subsection{Regional difference in performance} \label{methods_performance_loc}
\begin{figure}[ht]
        \centering
        \includegraphics[width=12cm]{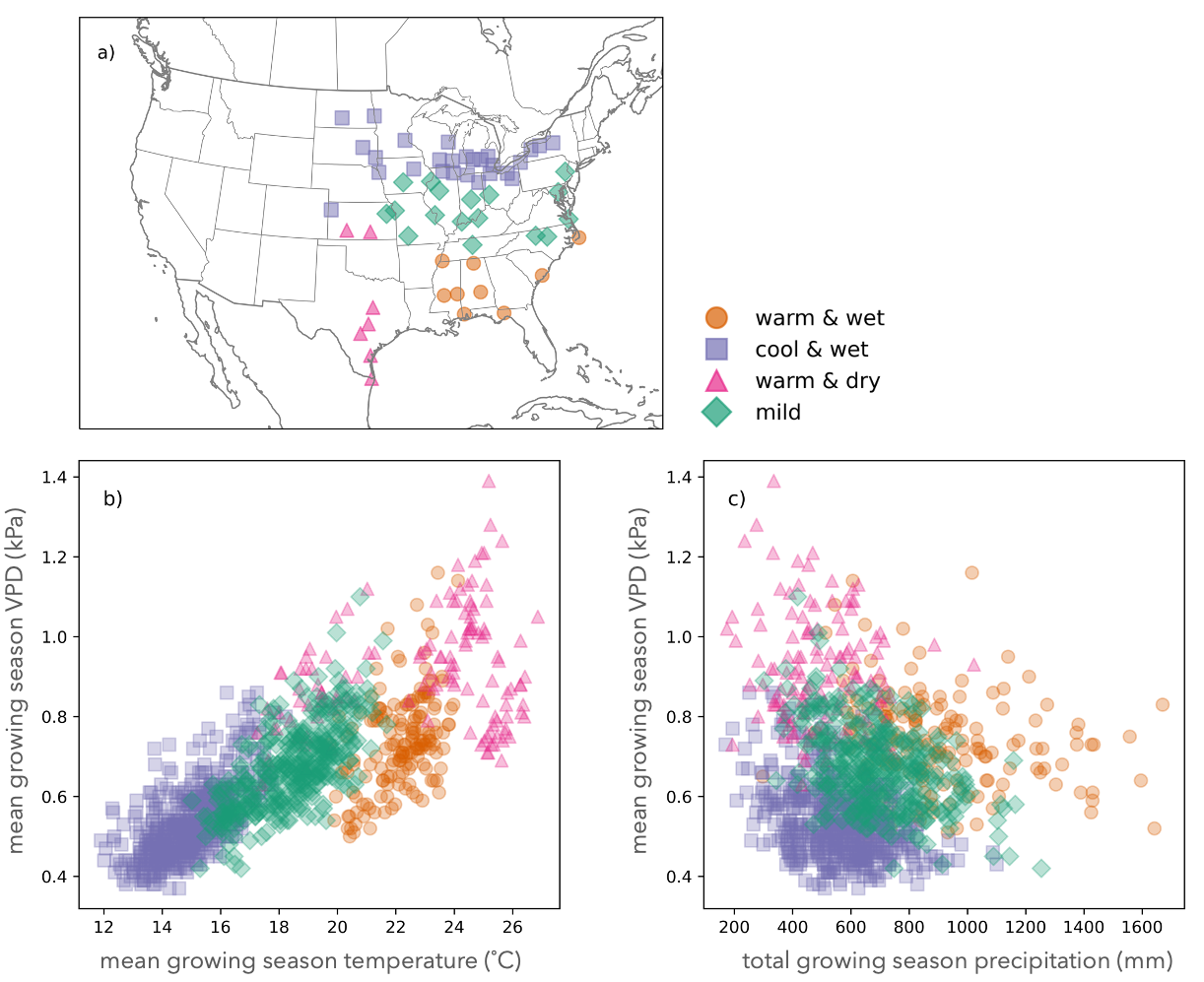}
        \caption{a) Map of simulation sites clustered based on mean growing season temperature (\textdegree{C}), mean growing season VPD (kPa), and total growing season precipitation (mm). Mean growing season temperature (\textdegree{C}), mean growing season VPD (kPa), and total growing season precipitation (mm) levels for all simulated site-years (b, c).}
        \label{fig:cspace_climate}
\end{figure}

We used a climate space approach to identify how the performance of T $\times$ M combinations differed with baseline climate conditions. We used a k-means clustering algorithm to cluster our sites based on mean growing season temperature and VPD, and total growing season precipitation, roughly dividing our simulation sites into four groups of climate spaces - cool and medium precip, mild, warm and wet, and warm and mild to dry (Fig. \ref{fig:cspace_climate}). We analyzed the performance of T $\times$ M combinations within each cluster of sites by calculating a standardized performance score (Eqn. \ref{eq:std_performance_score}):

\begin{equation}\label{eq:std_performance_score}
P_{score} = \frac{\sum_{i=1}^{n} R_{i}}{R_{max} * n}
\end{equation}

\noindent $R_{i}$ denotes the performance ranking of a T $\times$ M combination at site $i$ among a total of $n$ sites within each climate space, and $R_{max}$ indicates the maximum performance ranking a T $\times$ M combination can achieve at a single site. In our workflow, we only considered the top 20 performing T $\times$ M combinations when ranking high performing combinations (see section \ref{methods_performance}), so $R_{max}$ equals 20. The resulting standardized performance score ranges between 0 and 1, in which a T $\times$ M combination that ranks with highest in performance across all locations within the climate space would receive a $P_{score}$ of 1.

\subsection{In-season model outputs} \label{methods_inseason}
MAIZSIM generates outputs of a number of plant growth outputs throughout the growing season on an hourly time step. We describe in Table \ref{tab:maizsim_output} a select few outputs in more detail. We summarized these high time frequency outputs across four phenological stages (emerged, tassel initiation, tasseled and silked, and grain-filling) to facilitate analysis and interpretation. Specifically, we queried net photosynthetic rate ($A\textsubscript{n}$), net carbon gain ($P\textsubscript{n}$), and stomatal conductance ($g\textsubscript{s}$) values from daylight hours, and averaged them within the designated phenological stages. We represented ear biomass, total biomass, and total leaf area ($ear\_biomass$, $total\_biomass$, $LA$) within each developmental stage with maximum values within each stage. Finally, we queried water supply, demand, and deficit ($ET\_supply$, $ET\_demand$, $water\_deficit$) values at noon and averaged all values within each phenological stage, and represented predawn leaf water potential ($\Psi$) with values queried at 5 am, and averaged the all values within each phenological stage. 


\begin{table}[ht]
        \centering
        \caption{Key model outputs}
        \label{tab:maizsim_output}
        \begin{tabular}{lll}
        \hline
        Output & Description & Unit \\
        \hline
        $A_{n}$ & Net photosynthetic rate & µmol CO\textsubscript{2} m\textsuperscript{-2} sec\textsuperscript{-1} \\
        $P_{n}$ & Net carbon gain & g /plant hour \\
        $g_{s}$ & Stomatal conductance & g H\textsubscript{2}O m\textsuperscript{-2} sec\textsuperscript{-1} \\
        $ear\_biomass$ & Total ear biomass & g/plant \\
        $total\_biomass$ & Total plant biomass & g/plant  \\
        $LA$ & Total leaf area & cm\textsuperscript{2} \\
        $phenostage$ & Phenological stage & - \\
        $ET\_supply$ & Evapotranspiration (ET) supply & g H\textsubscript{2}O \\
        $ET\_demand$ & Evapotranspiration demand & g H\textsubscript{2}O \\
        $water\_deficit$ & ET demand - ET supply & g H\textsubscript{2}O \\
        $\Psi$ & Leaf water potential & MPa \\
        
        \hline
        \end{tabular}
\end{table}

\subsection{Experiment setup and model simulation}
\begin{figure}[htp]
        \centering
        \includegraphics[width=12cm]{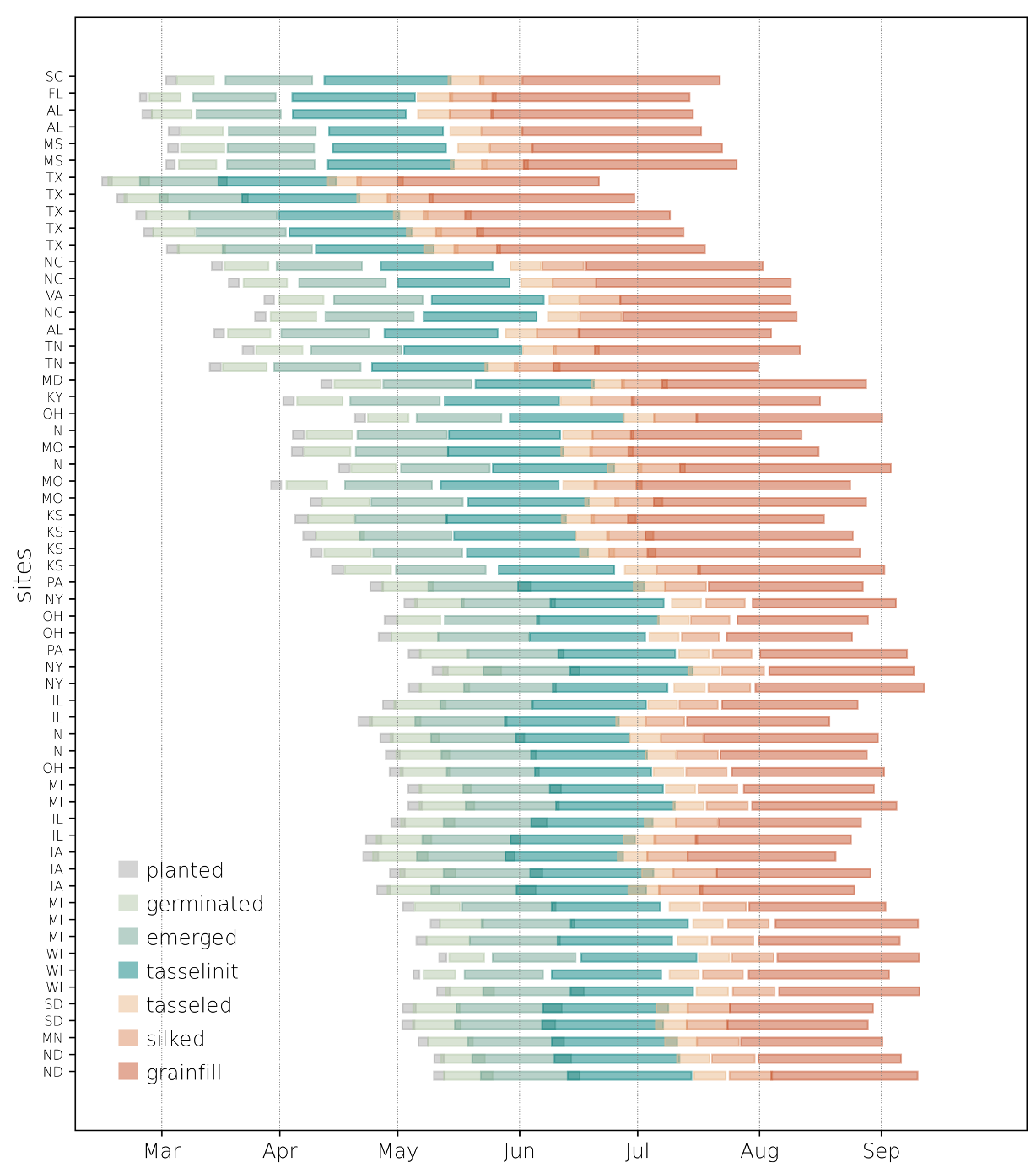}
        \caption{Start time and duration of each phenological stage across simulation sites, indicated by state abbreviations. Sites are roughly ranked by latitude, starting from southernmost sites towards the top.}
        \label{fig:bars_phenostage_sites}
\end{figure}

We prescribed the sampled planting density (Table \ref{table:params}, \emph{pop}) for each ensemble member and adjusted the planting date for each ensemble member and simulation site based on climate and growing degree days (GDD) requirements. We calculated GDD for each simulation site through accumulated heat units starting from February 1st with a base temperature of 8\textdegree C \citep{Kim_Modeling_2012} and determined the planting date once GDD surpassed the sampled for each ensemble member (Table \ref{table:params}, \emph{gdd}). This led to earlier planting dates in warmer regions and vice versa (Fig. \ref{fig:bars_phenostage_sites}), and created diversity in cropping cycle start time among T $\times$ M combinations, mimicking early versus late-planting cultivars (Fig. \ref{fig:bars_phenostage_phenos}). To simulate well-fertilized conditions, we prescribed a total of 200 kg ha\textsuperscript{-1} of nitrogen throughout the growing season, applying half as base fertilizer prior to planting and the rest top-dressed one month post planting.

For each simulation site, we ran the MAIZSIM model with default parameters that represented a generic crop cultivar across all locations (see default values in Table \ref{table:params}). Next, we carried out a site-level ensemble of simulations in which we used past meteorology data (see section \ref{methods_present_climate}) to each of the 100 trait-management parameter combinations (see section \ref{methods_params}) for each of our 1160 site--years (see section \ref{methods_performance}) and identified top performing (high yield and yield stability) trait-management combinations. Finally, we repeated the site-specific trait and management ensemble of simulations with idealized future climate (see section \ref{methods_future_climate}) to understand how high performing trait and management combinations under current climate conditions fared under future climate.

\subsection{Model validation}
\begin{figure}[htp]
        \centering
        \includegraphics[width=8cm]{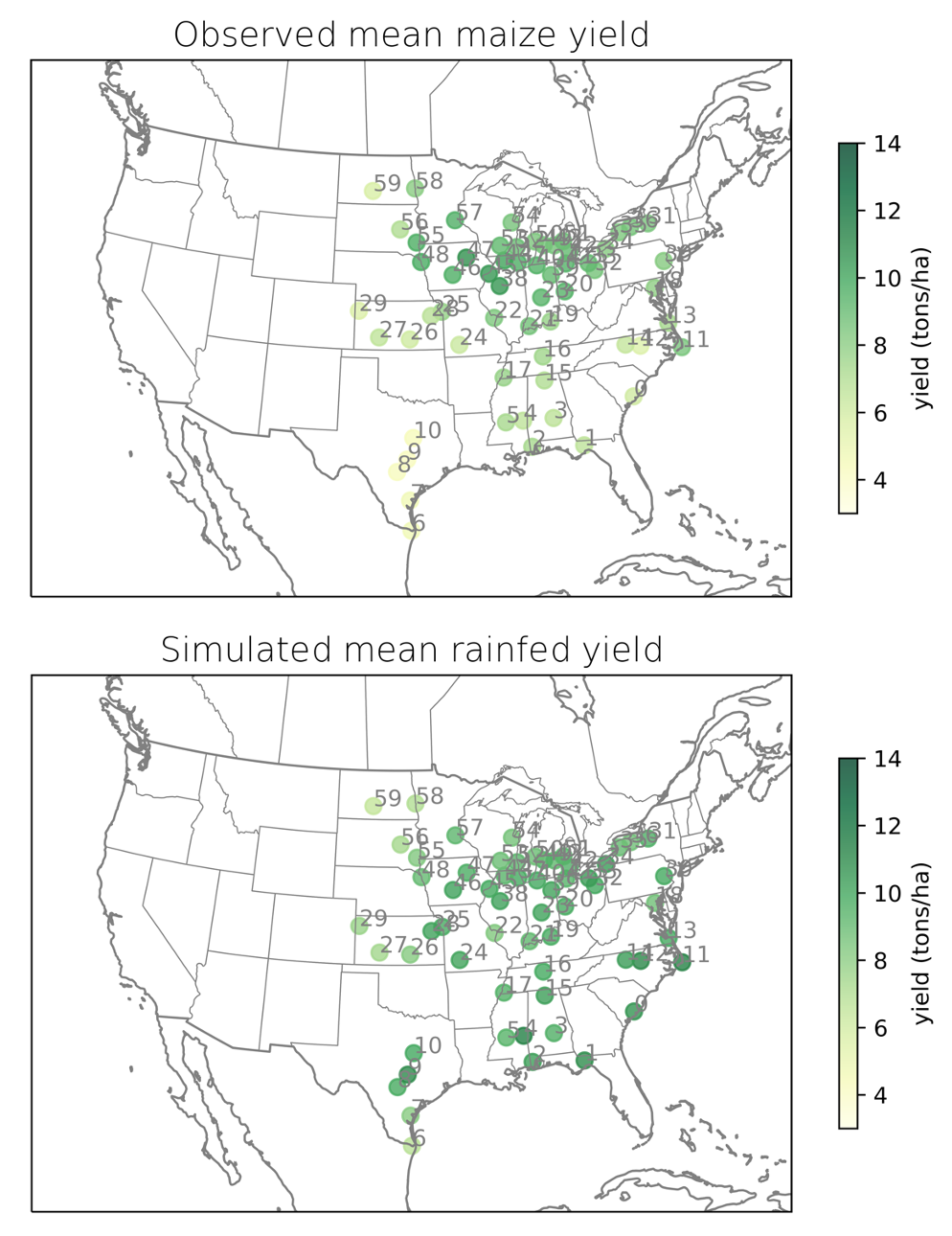}
        \caption{Observed (top) and simulated (bottom) yield (tons/ha) across simulated sites. Numbers indicate site numbers that correspond in Fig. \ref{fig:scatter_validate}.}
        \label{fig:maps_validate_sameloc}
\end{figure}

We validated simulated yields with default parameter values (control T $\times$ M combination, see Table \ref{table:params}) with historic yield data from the United States Department of Agriculture – National Agriculture Statistics Service (USDA- NASS, \url{https://www.nass.usda.gov/Data_and_Statistics/index.php}). We compared yield data from observation sites closest to our simulation sites calculated through a Euclidean distance (Fig. \ref{fig:maps_validate_sameloc}). We scaled our whole-plant level simulation outputs to field level by applying a planting density of 10 plants per m\textsuperscript{2} and compared our simulated yield with averaged yield observations in between years 2005-2012, since our default parameter and management options resemble modern-day plant traits, planting density, and planting dates.

\section{Results and discussion}
\subsection{Model validation}
\begin{figure}[ht]
        \centering
        \includegraphics[width=8cm]{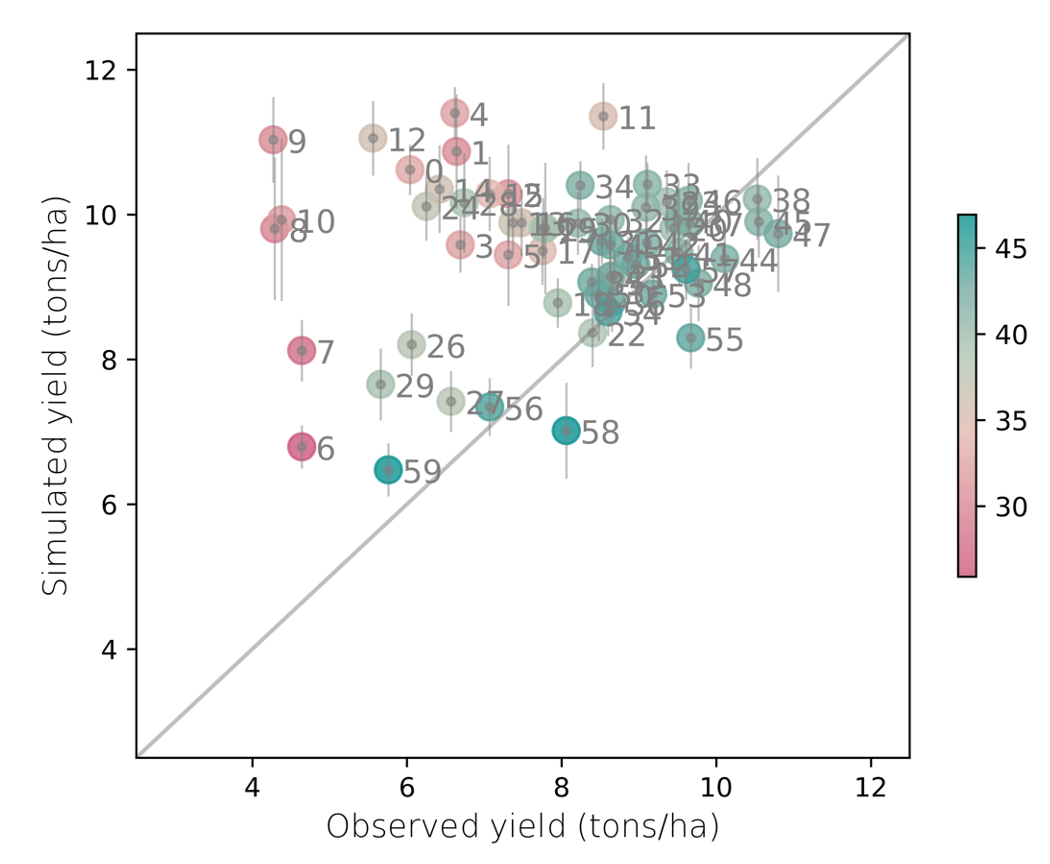}
        \caption{Observed versus simulated yield (tons/ha) across simulated sites. Colors indicate latitude of simulation site. Numbers correspond to site numbers shown in Fig. \ref{fig:maps_validate_sameloc}.}
        \label{fig:scatter_validate}
\end{figure}

In general, simulated yields showed less spatial difference compared to observations (Fig. \ref{fig:maps_validate_sameloc}). The model captured historic yield observations well in the higher latitude Corn Belt regions but overestimated yield in various warmer sites in southern locations (Fig. \ref{fig:scatter_validate}). Southern locations experience much warmer temperatures, especially during later parts of the growing season (Fig. \ref{fig:heatmap_temp}, grain-fill). While MAIZSIM dynamically describes temperature and water stress throughout the growing season through impacts on gas exchange and leaf developmental processes, the model lacks direct depiction of climate stress responses on reproduction processes such as flowering, pollination, and grain-filling, which are likely reasons for the yield over-estimation in warmer regions.  

Cultivar differences between crops planted in southern versus northern locations could also contribute to these discrepancies. Farmers in warmer southern locations choose cultivars that are both planted and harvested earlier in the growing season, leading to an overall shorter crop cycle duration compared to those planted in the north (USDA-NASS, Crop Progress and Conditions). While our simulation set up captures earlier planting in warmer regions (Fig. \ref{fig:bars_phenostage_sites}), it does not capture potential differences in cultivar and management choices that growers in the south have likely been opting for in order to avoid late season heat and water stress. Finally, we note that by applying a universal soil depth (200 cm), we may be overestimating soil water availability. This could disproportionately affect warmer locations in south, in which late-season water availability could partially alleviate water stress later in the growing season and contribute to yield overestimation in those locations (Fig. \ref{fig:scatter_validate}).

\subsection{Performance difference across climate spaces}
In \citeauthor{Hsiao2022} (\citeyear{Hsiao2022}, submitted), we followed the framework described in this paper and identified several top-performing strategies among all T $\times$ M combinations under present-day and future climate conditions, categorized based on different combinations of phenological (grain-filling start time and duration) and morphological (total leaf area) features. Top-performing strategies under present-day climate conditions included T $\times$ M combinations that either reached reproductive stage early (\emph{Early Starting}), were slow in aging (\emph{Slow Aging}), stress averting (\emph{Stress Averting}), or had large canopies and were high in yielding (\emph{High Yielding}). More details are described in Hsiao et al. (2022) and briefly summarized below. 

\begin{figure}[htp]
        \centering
        \includegraphics[width=12cm]{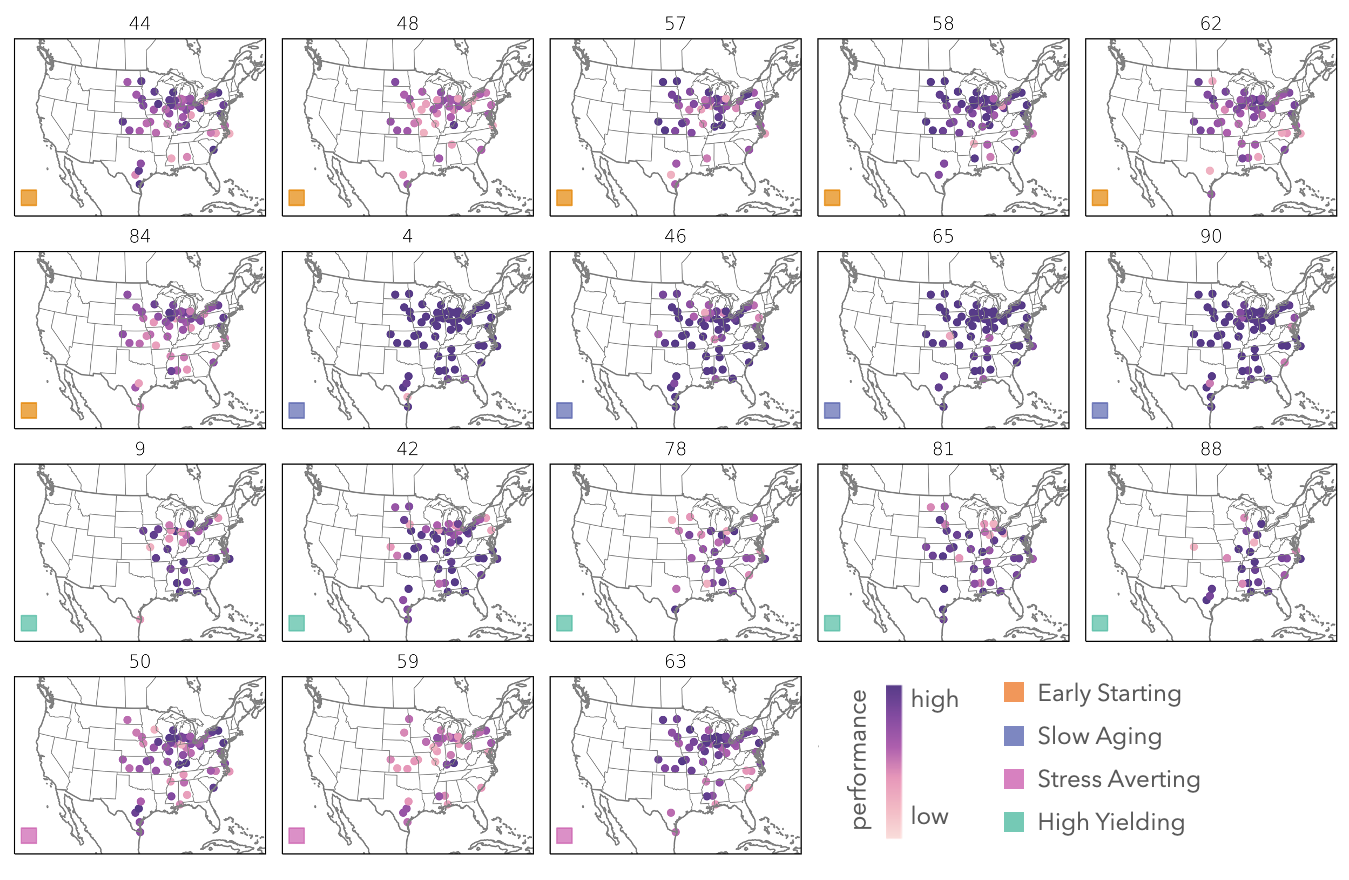}
        \caption{Performance ranking across simulation sites for T $\times$ M combinations of different top-performing strategies under present-day climate conditions.}
        \label{fig:rank_maps}
\end{figure}

\begin{figure}[htp]
        \centering
        \includegraphics[width=12cm]{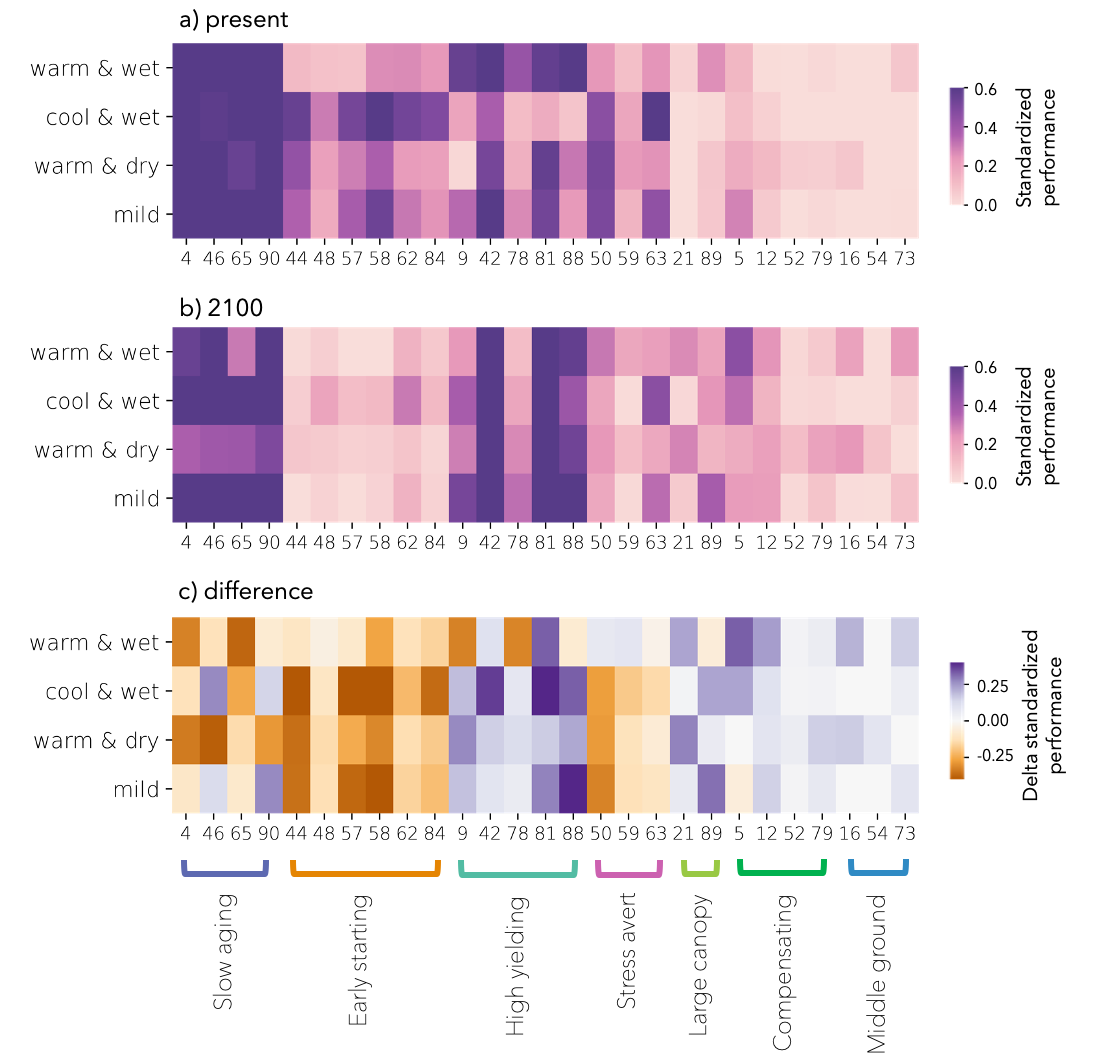}
        \caption{Standardized performance rankings of different strategies across climate spaces (see section \ref{methods_performance_loc}) under a) current, b) future climate conditions, and c) the difference between the two.}
        \label{fig:performance_cspace}
\end{figure}

\emph{Early Starting}, \emph{Slow Aging}, and \emph{Stress Averting} strategies all have a smaller canopy size and relatively earlier transition times from vegetative to reproductive stages, but differ in grain-filling duration. \emph{Slow Aging} strategies have long grain-filling durations that prolong cropping duration, while \emph{Stress Averting} strategies display the shortest longevity, allowing plants with this strategy to complete their cropping cycle early and avoid late season stressors such as dry and hot conditions. On the other hand, \emph{High Yielding} strategies have larger canopy sizes accompanied by delayed transition from vegetative to reproductive stages. While all categorized as top-performing under present-day climate, these strategies showcase a range of trade-offs between yield and yield stability, with performance differing across simulation sites (Fig. \ref{fig:performance}, \ref{fig:rank_maps}) and climate spaces (Fig. \ref{fig:performance_cspace}a).

\begin{figure}[htp]
        \centering
        \includegraphics[width=11cm]{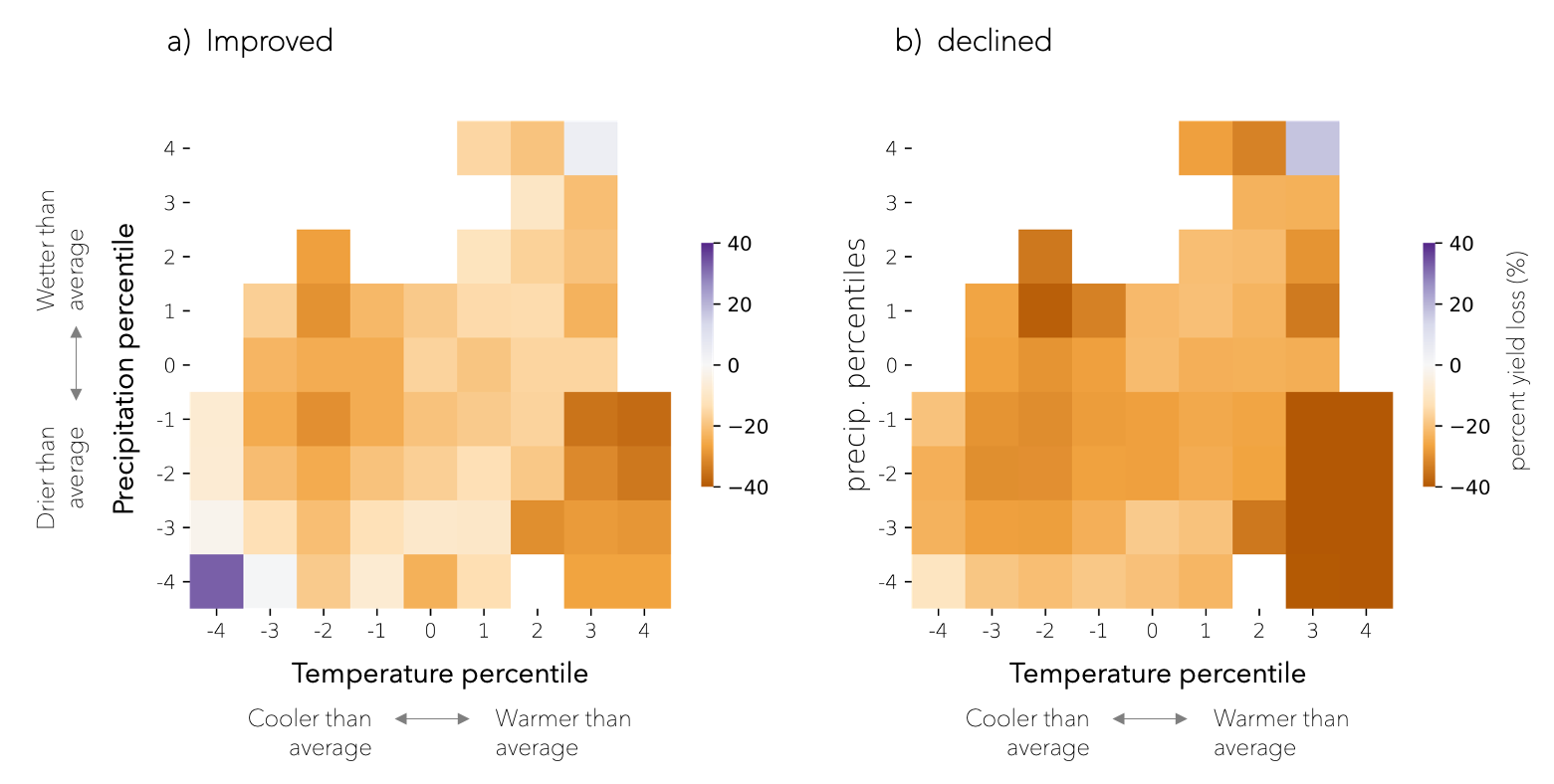}
        \caption{Percent yield loss within mean growing season temperature-precipitation climate space among T $\times$ M combinations that a) improved versus, b) declined in performance ranking under future climate conditions.}
        \label{fig:cspace_yieldloss_temp_precip}
\end{figure}

Under current climate conditions, T $\times$ M combinations with \emph{Slow Aging} strategies tend to be generalists, showing high performance across all climate spaces. On the other hand, strategies such as \emph{Early Starting} fared best in cool and wet regions, while \emph{High Yielding} strategies perform best under warm and wet conditions (Fig. \ref{fig:performance_cspace}a). Under future climate conditions, we observed an overall yield loss for all T $\times$ M combinations in most simulation sites (Fig. \ref{fig:cspace_yieldloss_temp_precip}), including strategies that improved in performance ranking with climate change (Fig. \ref{fig:cspace_yieldloss_temp_precip}a), such as \emph{High Yielding} and \emph{Large Canopy} (Fig. \ref{fig:performance_cspace}c). In general, warmer regions with low precipitation levels exhibited the greatest yield sensitivity (\% yield loss per degree C of warming, Fig. \ref{fig:cspace_yieldloss_temp_precip}); high-performing strategies under future climate partially buffered, but did not prevent yield loss. 

High-performing strategies under present-day climate conditions shifted with climate change (Fig. \ref{fig:performance_cspace}b, c). T $\times$ M combinations with early starting strategies experienced the greatest drop in performance ranking overall, showing declines in most climate spaces (Fig. \ref{fig:performance_cspace}c, \emph{Early Starting}). Slow aging strategies still remained one of the higher performers by the end of the century (Fig. \ref{fig:performance_cspace}b, \emph{Slow Aging}), but showed clear performance ranking declines in warm climate spaces (Fig. \ref{fig:performance_cspace}c, \emph{Slow Aging}), allowing several other strategies to compete for top performance in those climate spaces (Fig. \ref{fig:performance_cspace}c); T $\times$ M combinations with high yielding and compensating strategies progressed further in performance ranking (Fig. \ref{fig:performance_cspace}c, \emph{High Yielding}), and new high-performing strategies with larger canopies and delayed transition timings into reproductive stages emerged
(Fig. \ref{fig:performance_cspace}c, \emph{Large Canopy}, \emph{Compensating}, \emph{Middle Ground}).

\begin{figure}[htp]
        \centering
        \includegraphics[width=12cm]{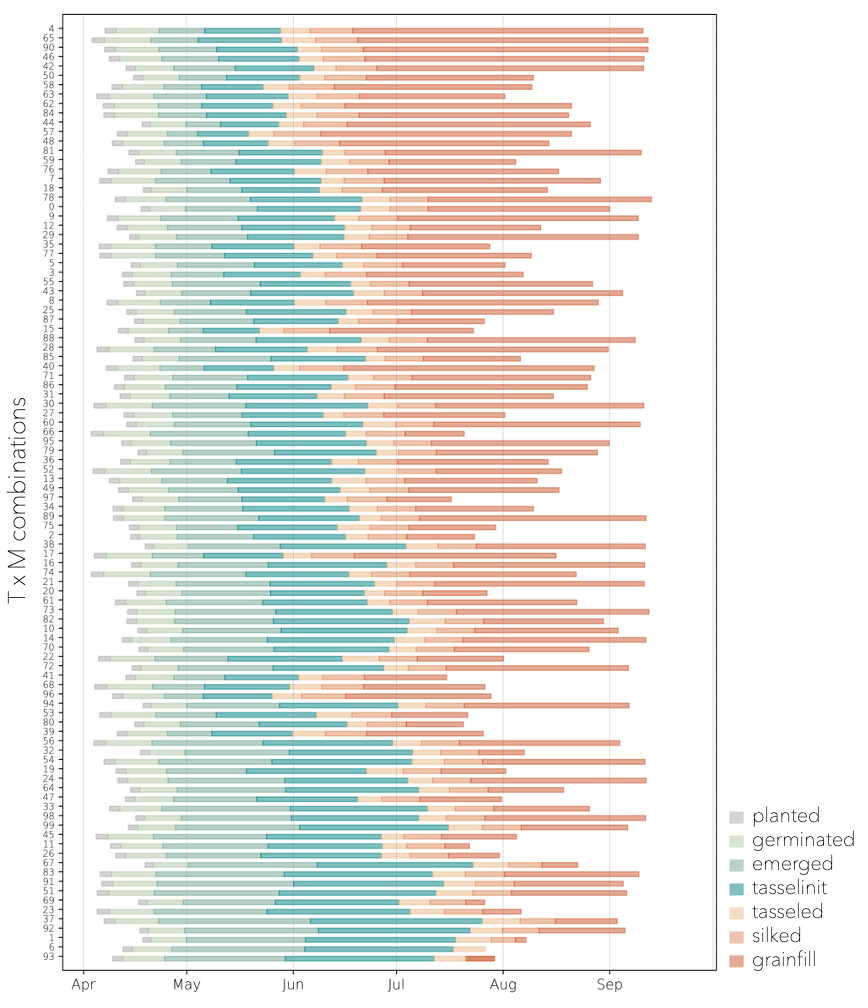}
        \caption{Start time and duration of each phenological stage across T $\times$ M combinations, averaged across all simulation sites, ranked by overall performance, with the highest performers listed towards the top.}
        \label{fig:bars_phenostage_phenos}
\end{figure}

\begin{figure}[htp]
        \centering
        \includegraphics[width=12cm]{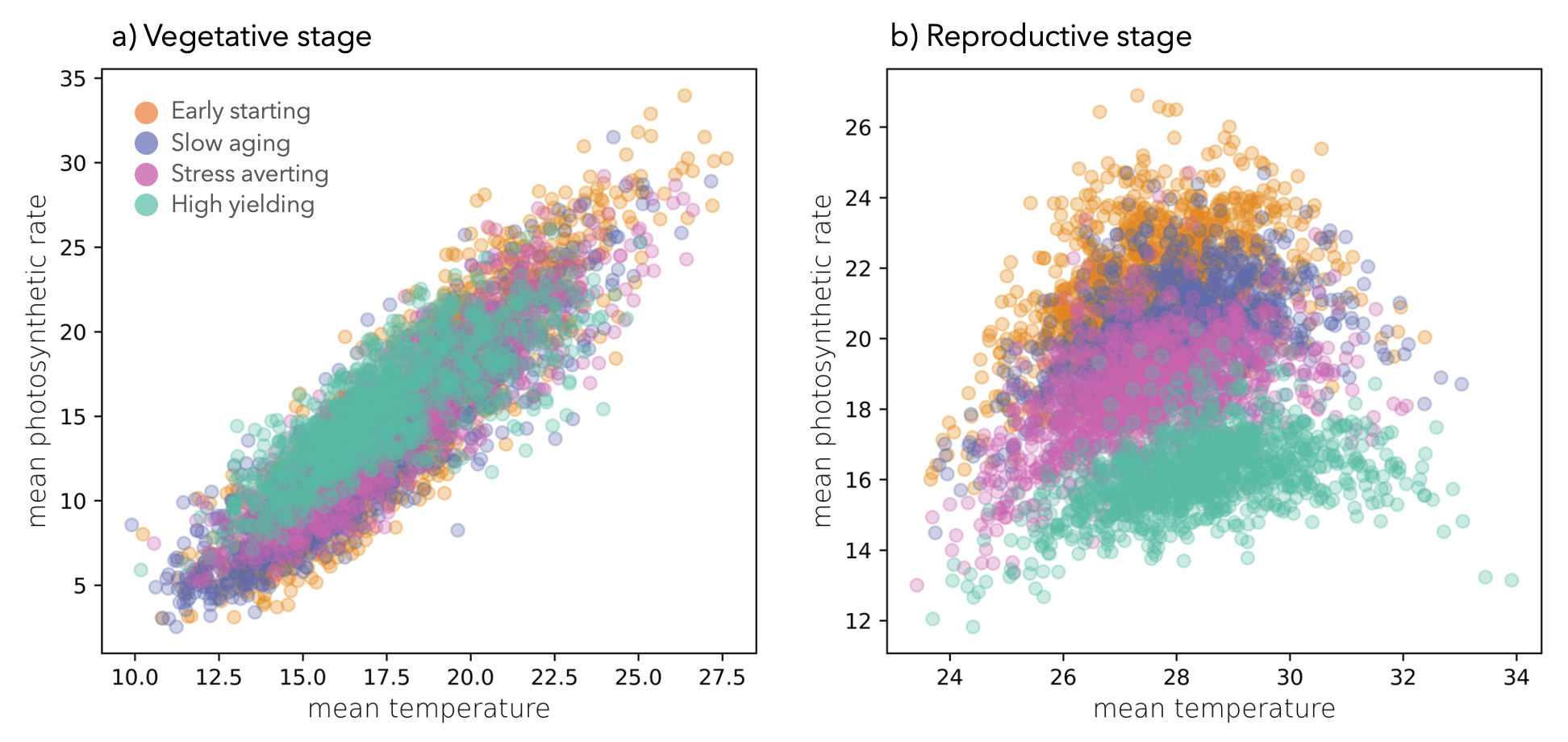}
        \caption{Relationship between mean temperature (\textdegree{C}) and mean photosynthetic rate (µmol CO\textsubscript{2} m\textsuperscript{-2} sec\textsuperscript{-1}) for all simulated site-years during a) vegetative versus b) reproductive stages for four representative T $\times$ M combinations within top-performing strategies.}
        \label{fig:scatter_temp_photo}
\end{figure}

\subsection{Mechanisms for high performance}
We analyzed detailed in-season outputs of various phenological, physiological, and morphological outputs of the model (Table \ref{tab:maizsim_output}) and describe here some general trends observed in top-performing T $\times$ M combinations. 

\subsubsection{Phenology}
\begin{figure}[ht]
        \centering
        \includegraphics[width=12cm]{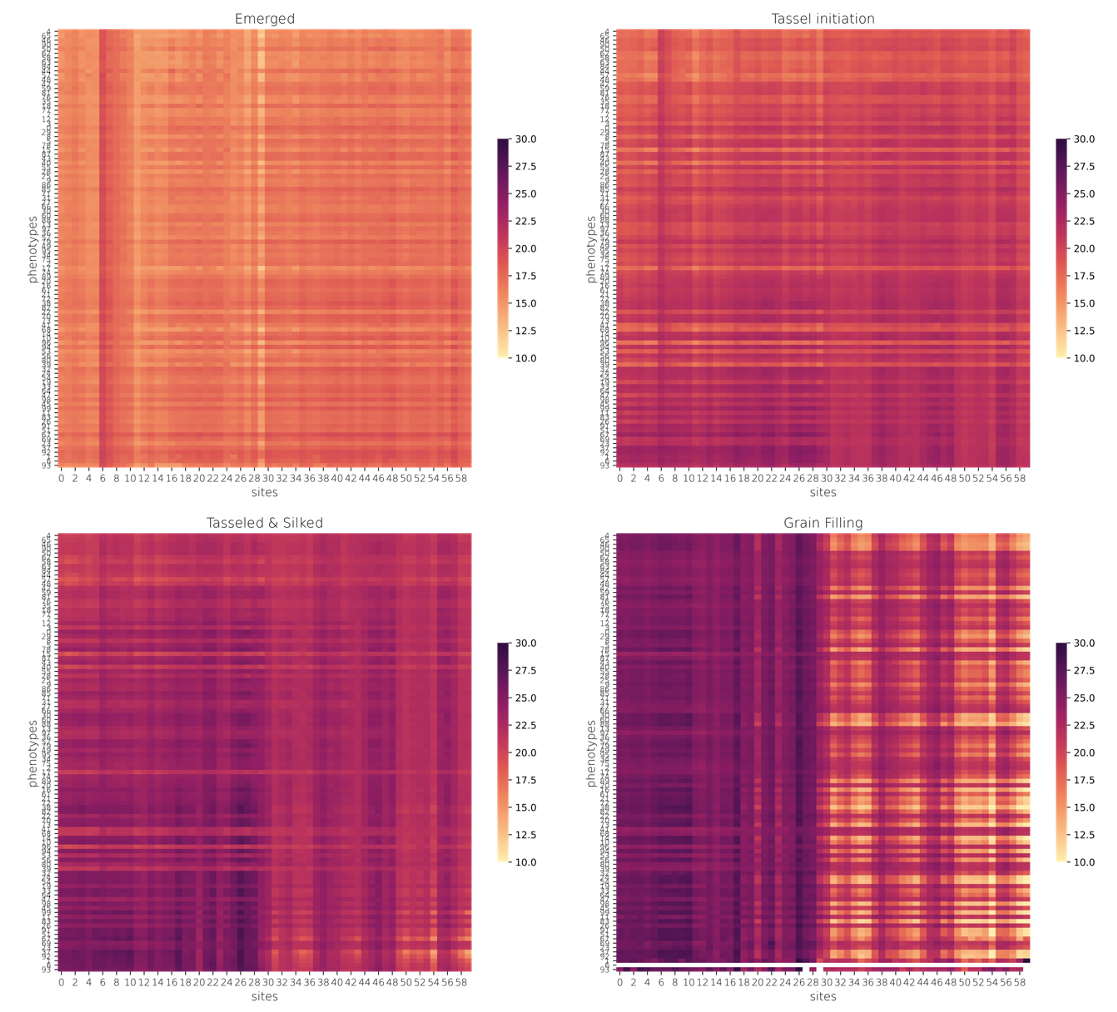}
        \caption{Mean air temperature (\textdegree{C}) across phenological stages for top phenotypes across all sites, ranked by performance of T $\times$ M combinations, and averaged within phenological stages.}
        \label{fig:heatmap_temp}
\end{figure}

\begin{figure}[htp]
        \centering
        \includegraphics[width=12cm]{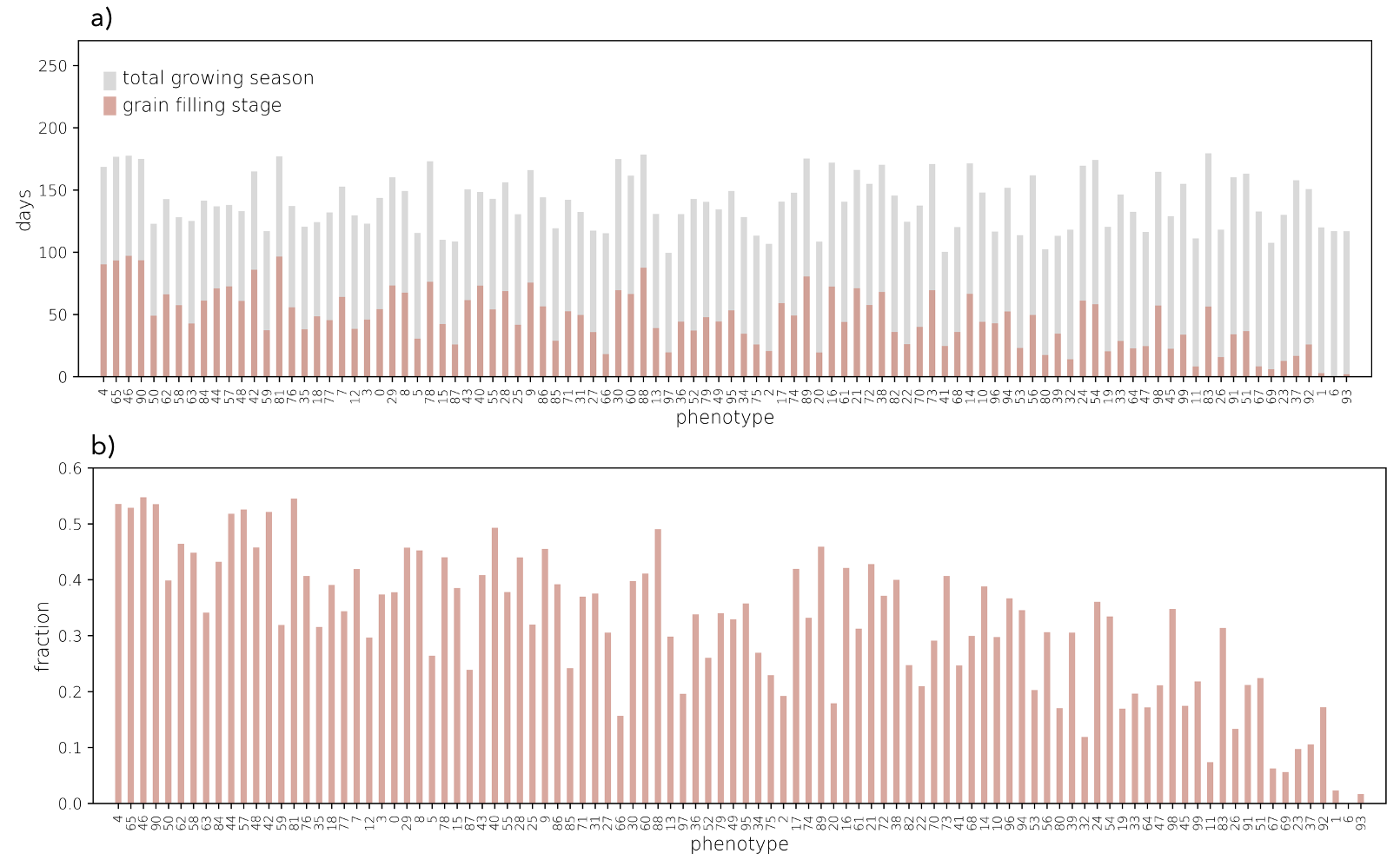}
        \caption{a) Total growing season length (gray, days) and grain-filling length (pink, days) for all phenotypes, ranked by performance, starting with the highest performers towards the left, and b) the fraction of grain-filling over total growing season.}
        \label{fig:bars_gflen_frac}
\end{figure}

Climatological differences between simulation sites and parameter differences among T $\times$ M combinations both lead to the range of phenology output we see in our simulation outputs (Fig. \ref{fig:bars_phenostage_sites}, \ref{fig:bars_phenostage_phenos}). Phenology differs across simulation sites due to imposed planting date adjustments based on growing degree day requirements, allowing for an earlier planting date in warmer regions (Fig. \ref{fig:bars_phenostage_sites}). Climatological differences throughout the growing season further shape the difference, especially during the grain-filling stage in which simulation sites in the south become substantially warmer than those in the north (Fig. \ref{fig:heatmap_temp}), leading to hastened development (Fig. \ref{fig:bars_phenostage_ns}). On the other hand, phenology differs among T $\times$ M combinations due to differences in perturbed traits linked to phenology (e.g. planting time, developmental rate, leaf number, Fig. \ref{fig:bars_phenostage_phenos}). High performing T $\times$ M combinations under present-day climate conditions tend to show earlier starts in reproductive stages with a longer duration (Fig. \ref{fig:bars_phenostage_phenos}). Higher ranking T $\times$ M combinations tend to show a greater fraction of grain-filling length over total growing season length (Fig. \ref{fig:bars_gflen_frac}b) despite no clear trends in total growing season length (Fig. \ref{fig:bars_gflen_frac}a).

\subsubsection{Physiology}
\begin{figure}[htp]
        \centering
        \includegraphics[width=12cm]{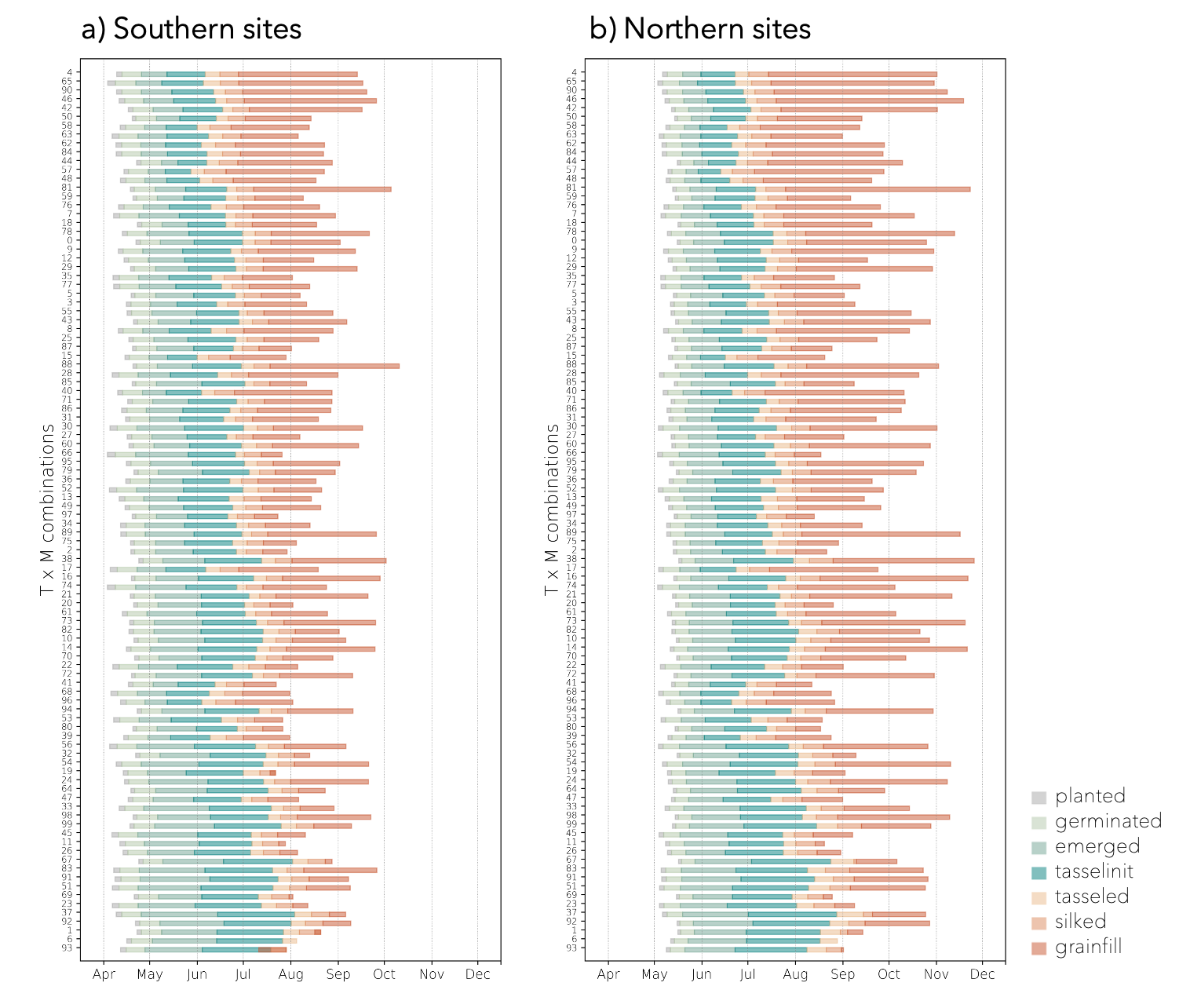}
        \caption{Start time and duration of each phenological stage across T $\times$ M combinations, averaged across all simulation sites for a) southern sites versus b) northern sites, ranked by overall performance, with the highest performers listed towards the top.}
        \label{fig:bars_phenostage_ns}
\end{figure}

\begin{figure}[htp]
        \centering
        \includegraphics[width=12cm]{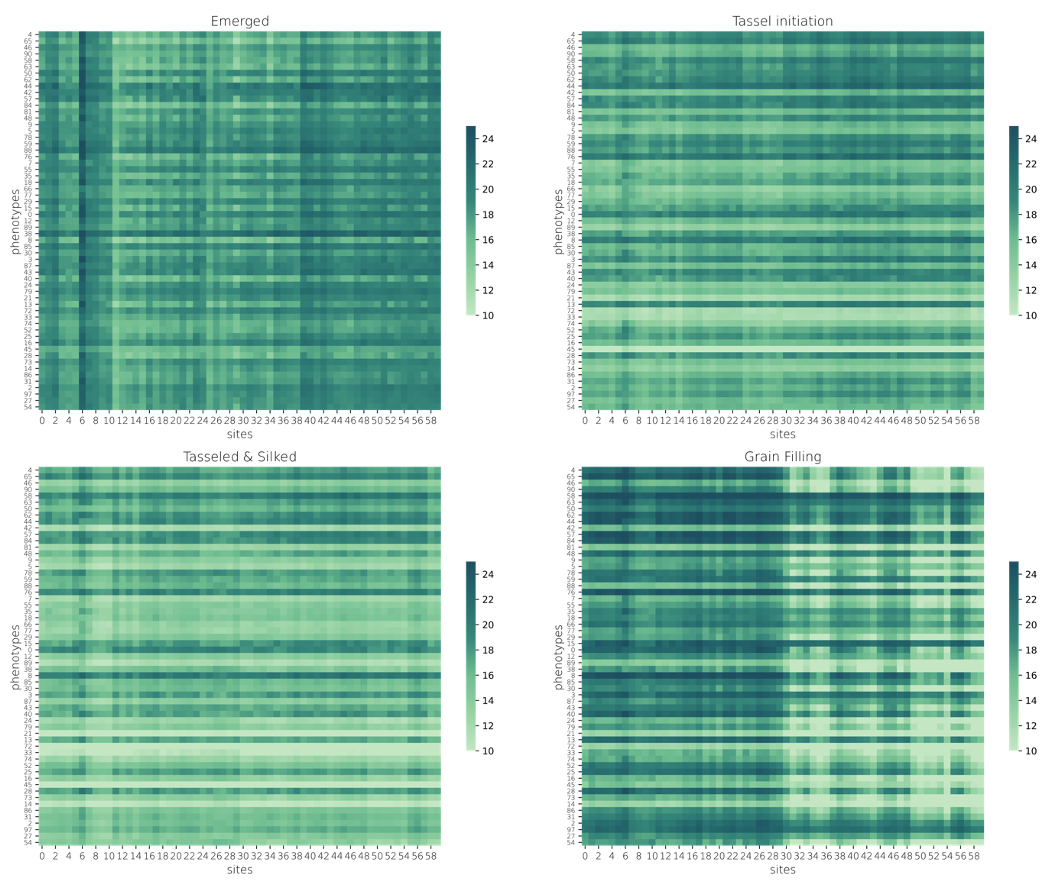}
        \caption{Net photosynthetic rate (µmol CO\textsubscript{2} m\textsuperscript{-2} sec\textsuperscript{-1}) across phenotypes and sites, ranked by the performance of T $\times$ M combinations, and averaged within each developmental stage.}
        \label{fig:heatmap_An}
\end{figure}

\begin{figure}[htp]
        \centering
        \includegraphics[width=12cm]{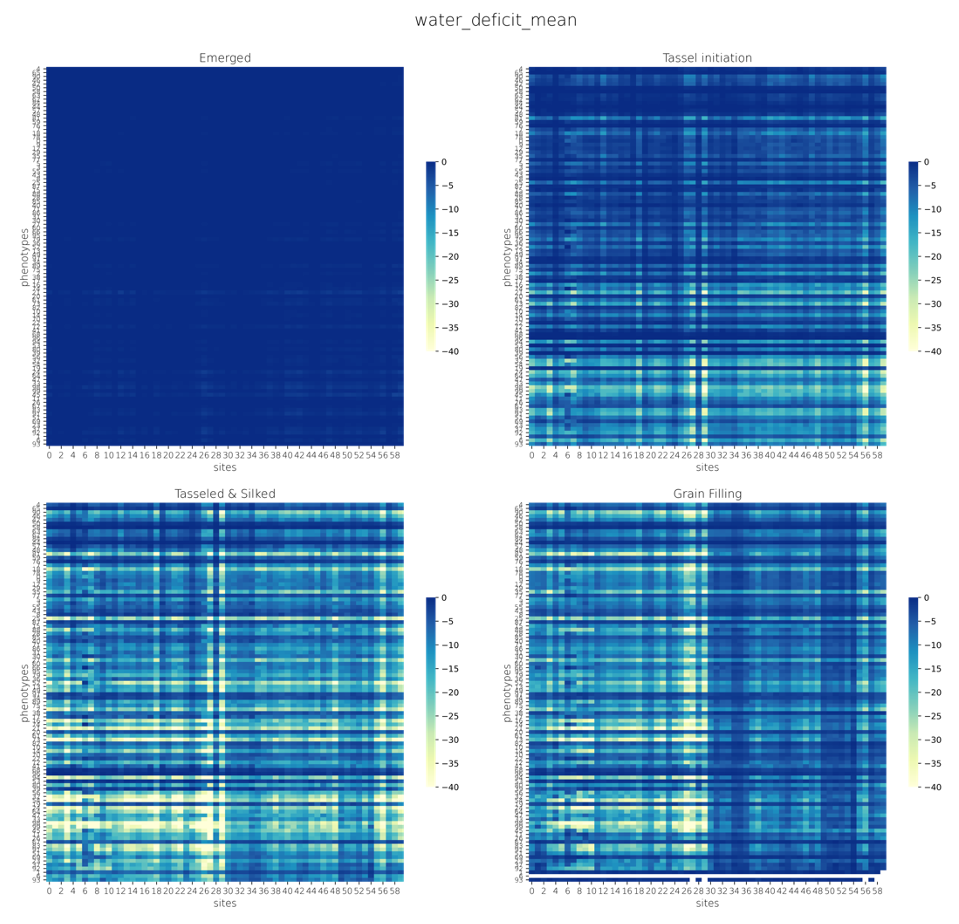}
        \caption{Mean water deficit (g H\textsubscript{2}O) across phenotypes and sites, ranked by the performance of T $\times$ M combinations, and averaged within each developmental stage.}
        \label{fig:heatmap_WD}
\end{figure}

\begin{figure}[htp]
        \centering
        \includegraphics[width=12cm]{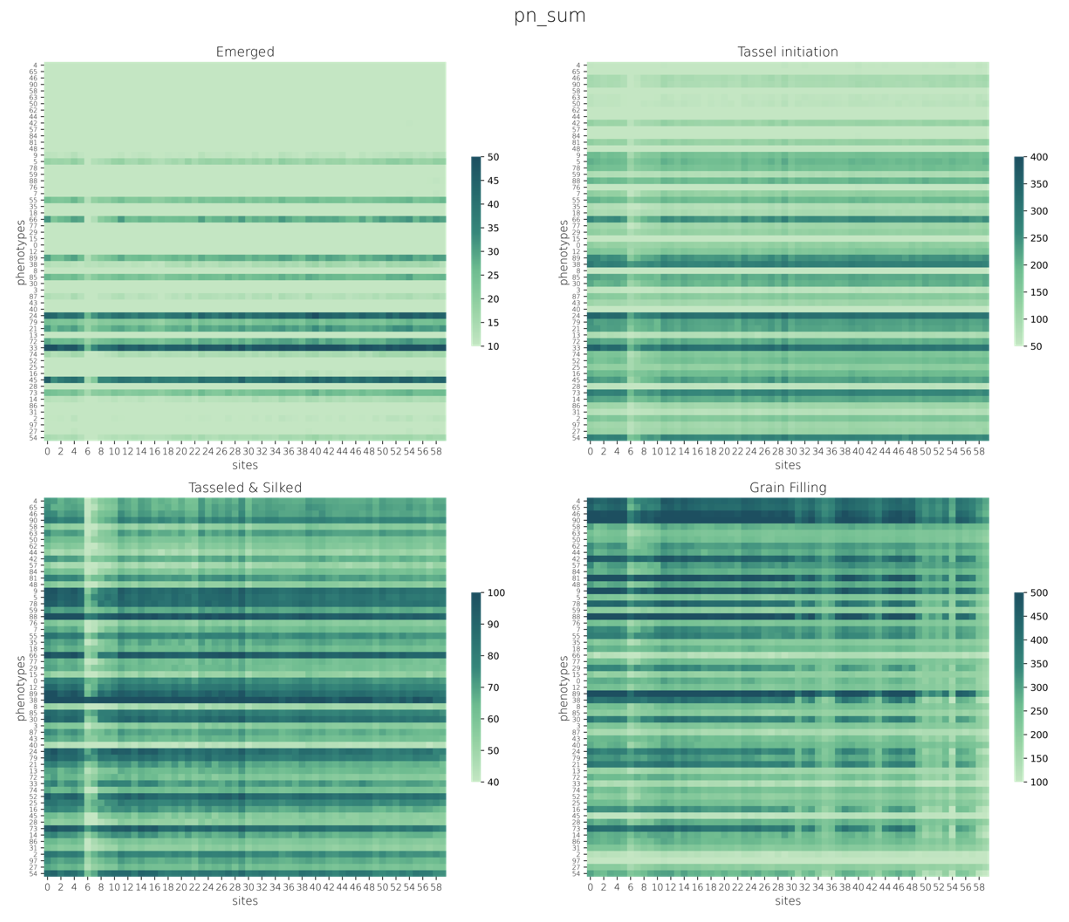}
        \caption{Net carbon gain throughout phenological stage (g C) across phenotypes and sites, ranked by the performance of T $\times$ M combinations, and averaged within each developmental stage.}
        \label{fig:heatmap_pn_sum}
\end{figure}

\begin{figure}[htp]
        \centering
        \includegraphics[width=12cm]{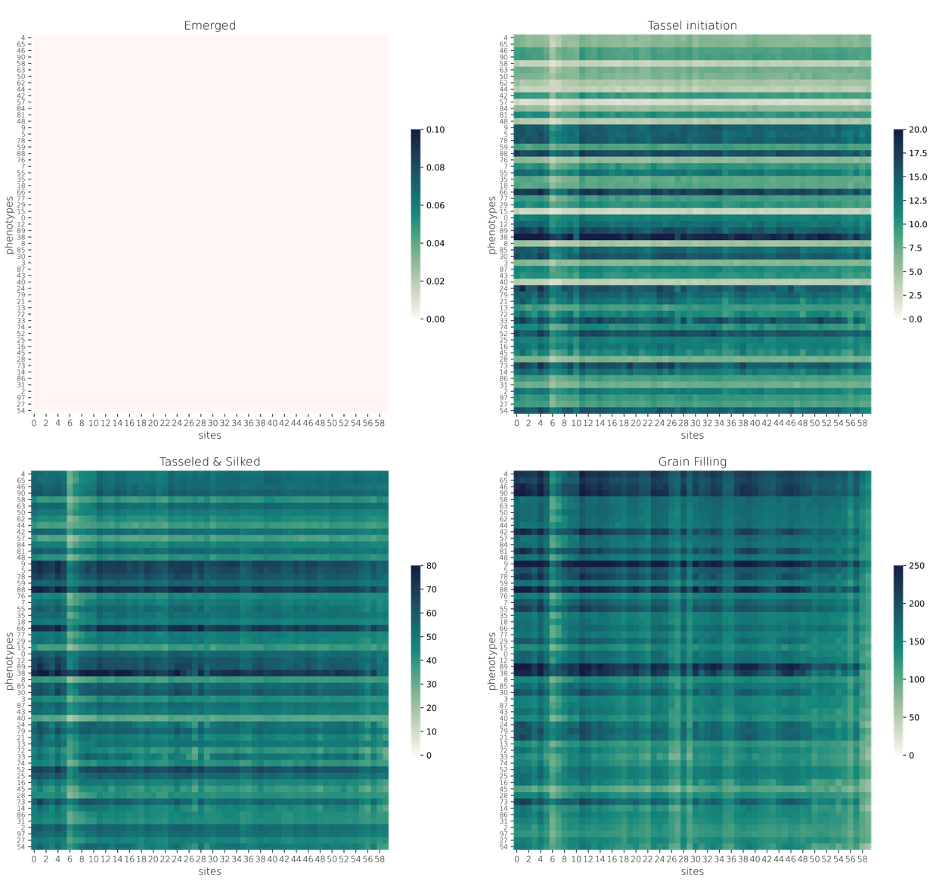}
        \caption{Ear biomass (g/plant) across phenotypes and sites, ranked by the performance of T $\times$ M combinations, and averaged within each developmental stage.}
        \label{fig:heatmap_dm_ear}
\end{figure}

Net photosynthetic rates are generally higher in top-performing T $\times$ M combinations during transition from vegetative into reproductive stages (Fig. \ref{fig:heatmap_An}), but the differences in photosynthetic rates become dominated by climatological differences between simulation sites during the final grain-filling stage, with greater photosynthetic rates in warmer southern regions (Fig. \ref{fig:heatmap_An}). In general, we see a linear relationship between temperature and photosynthetic rate during vegetative stages (Fig. \ref{fig:scatter_temp_photo}). This relationship eventually plateaus around 28-30 \textdegree{C} later in the growing season, and starts to decline in a few warmest site-years (Fig. \ref{fig:scatter_temp_photo}b). While warmer temperatures generally led to higher photosynthetic rates, hastened development and greater water deficit under warmer conditions also led to overall shorter grain-filing durations (Fig. \ref{fig:heatmap_WD}, \ref{fig:bars_phenostage_ns}a), compensating one another, dampening the overall difference between northern versus southern sites in terms of net carbon gained throughout grain-filling (Fig. \ref{fig:heatmap_pn_sum}) and final yield (Fig. \ref{fig:heatmap_dm_ear}).

\subsubsection{Morphology}
\begin{figure}[htp]
        \centering
        \includegraphics[width=12cm]{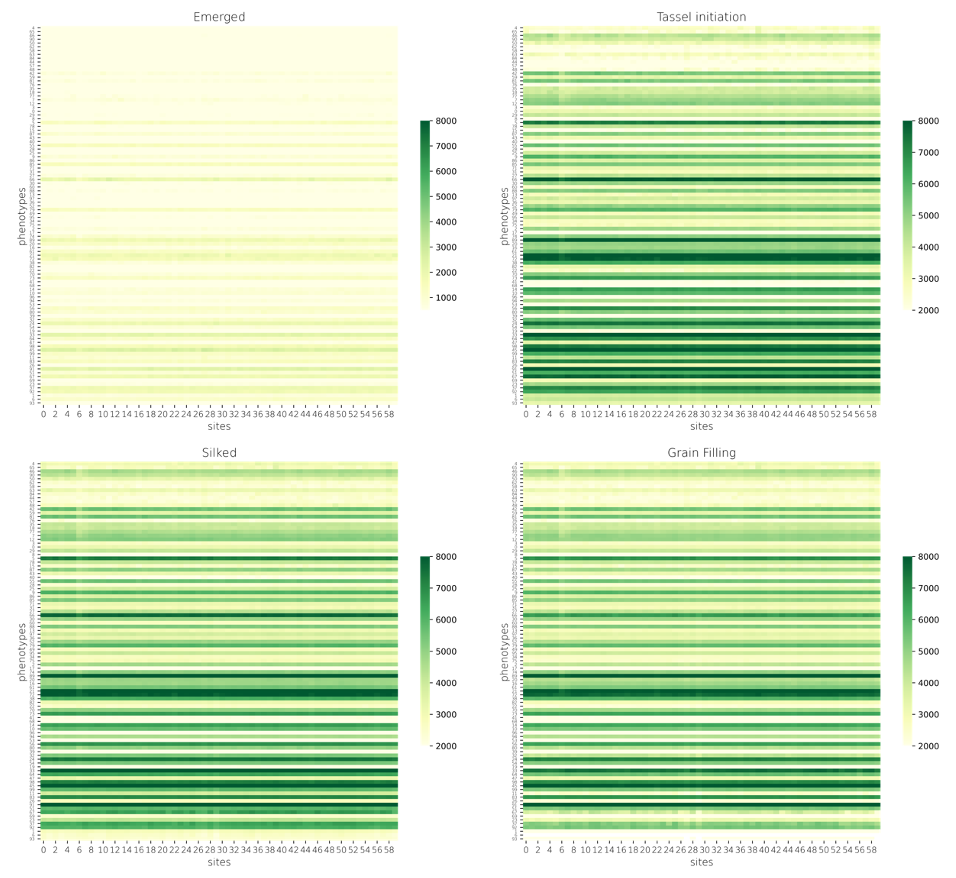}
        \caption{Total leaf area (cm\textsuperscript{2}) across phenotypes and sites, ranked by the performance of T $\times$ M combinations, and averaged within phenological stages..}
        \label{fig:heatmap_LA}
\end{figure}

Differences in simulated total leaf area was largely dominated by parameter make up within T $\times$ M combinations, showing much less difference in simulated yield across sites (Fig. \ref{fig:heatmap_LA}). Simulated plants approached full canopy sizes around tassel initiation, and top-performing combinations showed mid to lower total leaf area under present-day climate conditions. This was consistent with most top-performers under present-day climate exhibiting early transitions into reproductive stage (e.g., \emph{Slow Aging}, \emph{Early Starting}, \emph{Stress Averting}). These strategies partly achieved early reproductive start through short vegetative stages through fewer total number of leaves and hence smaller canopy size (i.e., lower total leaf area).

\section{Discussion}
Crop production is expected to suffer under future climate conditions as the climate warms. Adaptation of crop management practices, location of planting, as well as adaptation of the crops themselves all have the potential to limit expected yield loss and help to sustain agricultural productivity. However, we lack a systematic understanding of which adaptations are likely to have the biggest impact and why, both critical pieces of knowledge for agricultural planning. Mechanistic, process-based crop simulation models can be a powerful tool to synthesize cropping information, set breeding targets, and develop adaptation strategies for sustaining food production, yet have been underutilized for developing specific climate-adaptation options.      

Breeding for and adopting new cultivars involve exploring and navigating the hills and valleys of the G $\times$ E $\times$ M landscape, in which optimal plant traits and management options are identified within defined target environments \citep{Messina2011a, Cooper2016}. Requirements from breeding, delivering, and adopting a desirable cultivar depends on many factors, and the whole process can take from years to decades \citep{Challinor2017}. Recent developments in breeding practices have greatly expanded the efficiency in genotyping and phenotyping methods \citep{Voss-Fels2019}, yet the breeding pipeline is still time and resource intensive, limiting its ability to explore the full range of G $\times$ E $\times$ M combinations and interactions.

A typical breeding cycle starts out by exposing a large germplasm pool under extremely high selection pressure, filtering genotypes from the order of 10$^6$ individuals down to a few dozen promising candidates \citep{Messina2020, Cooper2014}. In the early stages of a breeding program, trait selection is often limited to those that can easily be identified through automated processes, and commonly based on plants in early developmental stages. It is not until later in the breeding cycle that selection criteria shift from genotype to phenotype-based, and promising hybrids are evaluated on-farm at various locations with a range of background climate conditions \citep{Gaffney2015}. Management optimization also occurs around this time, in which field trials are set up to identify the best management practices for the final candidates prior to commercial release. Further, common breeding methods that either select for higher yield or eliminate defect traits do not allow for a clear understanding of the mechanisms in which favorable traits contribute to greater performance and yield, and effective combinations of plant traits, if not actively sought for based on a mechanistic understanding of crop growth and yield, could only occur by chance \citep{Donald1968}. 

There is growing recognition that mechanistic crop simulation models can be a powerful tool to synthesize cropping information, set breeding targets, and develop adaptation strategies for sustaining food production. Such applications can complement current breeding efforts of developing new climate-resilient cultivars by
facilitating broad exploration of the G $\times$ E $\times$ M landscape within a modeled setting as a first step \citep{Muller2019,Rotter2015,Cooper2020,Messina2020b}. The process-based nature of such models allow for mechanistic insight through which these adaptations influence yield and their sensitivity to different climate factors, providing a more complete assessment of the uncertainty associated with different possible climate conditions, including those that do not currently exist yet. 

\citeauthor{Ramirez-Villegas2015} (\citeyear{Ramirez-Villegas2015}) provided a few successful examples of model-aided breeding projects, such as the New Plant Type program developed by IRRI for rice crops, in which process-based models were used to help make informed decisions to target breeding directions and resulted in new plant types that out-yielded conventional cultivars within two breeding cycles \citep{Peng2008}. This success further inspired the super rice program in China that led to newly developed rice varieties with 15-25\% higher yield than common hybrid cultivars planted in other regions in China \citep{Peng2008}. While model-aided breeding practices are less commonly targeted towards a changing climate \citep{Ramirez-Villegas2015}, demonstrated success under current climate suggest it as a promising pathway to guide breeding direction for climate adaptation moving forward, and expanded experiments evaluating a range of G $\times$ E $\times$ M conditions can enable production system responses to changing environmental conditions (Messina et al., 2020, Wang et al., 2019).

Regardless, thorough evaluation and application of crop models for developing specific climate-adaptation options (e.g., designing adaptive phenotypes for specific soil-climate combinations) for US agriculture remains scarce. We bridge this gap by constructing an integrated data-model framework set up to explore crop performance across a defined G $\times$ E $\times$ M landscape. With this framework, we identified high-performing plant trait and management combinations (G $\times$ M) best suited for current climate conditions, as well as targets and priorities to adapt to impending climate stressors (E). Heterogeneity in performance exists within the sampled climate space, which stemmed from underlying physiological, morphological, and phenological processes within the simulated crop. Model outputs on hourly time steps allowed us to compile detailed in-season outputs of various plant processes and summarize them according to associated phenological stages. This form of model output allows for more in-depth analysis that go beyond final yield and yield stability, and investigation of mechanisms that contribute to high crop performance and the differences across climate spaces and under future climate projections. 

We demonstrate how such a framework can be used to identify adaptation options with an emphasis on climate-resilient plant traits and effective management that will mitigate yield loss and optimize productivity both across space and through time in US corn growing regions under future climate conditions. Our modeling results demonstrate that application of mechanistic modeling holds substantial promise to inform breeding within the US maize production system.

\section{Acknowledgements}
We thank Lucas Vargas Zeppetello for providing CMIP6 temperature and relative humidity projection scaling patterns. JH acknowledges support from the AFRI NIFA Fellowships Grant Program: Predoctoral Fellowships [grant no. 2020-67034-31736] from the USDA National Institute of Food and Agriculture. ALSS acknowledges support from NSF AGS-1553715.

\pagebreak
\printbibliography

\pagebreak
\beginsupplement
\section{Supplementary information}

\end{document}